\documentclass[webpdf,traditional,mediumone]{oup-authoring-template}

\graphicspath{{Fig/}}

\theoremstyle{thmstyleone}%
\theoremstyle{thmstyletwo}%
\theoremstyle{thmstylethree}%

\usepackage{booktabs}
\usepackage{hyperref}
\hypersetup{colorlinks=true,linkcolor=blue,citecolor=blue,urlcolor=blue}
\usepackage{pdfpages}
\begin{document}

\journaltitle{Biostatistics}
\DOI{DOI added during production}
\copyrightyear{YEAR}
\pubyear{YEAR}
\vol{XX}
\issue{x}
\access{Published: Date added during production}
\appnotes{Paper}

\firstpage{1}


\title[Continuous-Time Bayesian Networks with Structured Shrinkage Priors for Modelling Multimorbidity Trajectories]{Continuous-Time Bayesian Networks with Structured Shrinkage Priors for Modelling Multimorbidity Trajectories in Large-Scale Electronic Health Records}

\author[1,$\ast$]{Oyebayo~R.~Olaniran\ORCID{0000-0001-7342-8639}}
\author[1]{Soumya~S.~Paria\ORCID{0009-0007-0631-4050}}
\author[2]{Mizanur~Khondoker\ORCID{0000-0002-1801-1635}}
\author[2]{Alexander~J.~MacGregor\ORCID{0000-0003-2163-2325}}
\author[1]{Alexandra~Lewin\ORCID{0000-0003-0081-7582}}

\address[1]{\orgdiv{Department of Medical Statistics}, \orgname{London School of Hygiene \&
Tropical Medicine}, \orgaddress{\street{Keppel St.}, \postcode{WC1E 7HT}, \state{London}, \country{United Kingdom}}}
\address[2]{\orgdiv{Norwich Medical School}, \orgname{University of East Anglia}, \orgaddress{\street{Norwich Research Park Colney Lane}, \postcode{NR4 7UJ}, \state{Norwich}, \country{United Kingdom}}}

\corresp[$\ast$]{Corresponding author. \href{email:email-id.com}{ridwan.olaniran@lshtm.ac.uk}}

\received{Date}{0}{Year}
\revised{Date}{0}{Year}
\accepted{Date}{0}{Year}



\abstract{
Multiple long-term conditions (MLTCs) arise through complex, time-dependent interactions among diseases, yet existing methods often struggle to jointly model disease progression, multimorbidity networks, and high-dimensional risk factors. We propose a structured Bayesian continuous-time Bayesian network (CTBN) framework for learning directed disease-dependency networks from longitudinal electronic health records. The model allows disease transition intensities to depend on existing conditions, pairwise disease interactions, and exogenous covariates. To control the combinatorial growth of interaction parameters, we introduce order-dependent shrinkage priors that increasingly penalise higher-order effects while preserving clinically interpretable main effects. We compare four sparsity-inducing priors, spike-and-slab, structured normal, Bayesian LASSO, and regularised horseshoe through extensive simulation studies. Across multiple data-generating scenarios, the spike-and-slab prior achieved the best network recovery, variable-selection accuracy, and false-discovery control, while continuous shrinkage priors were less effective for hard variable selection. The proposed framework was applied to UK Biobank primary care records, focusing on data from 33,558 participants who were free of the ten selected most prevalent conditions at age 40 and who subsequently developed at least one of these conditions during the follow-up period. The selected spike-and-slab model identified two dominant disease modules: a cardiometabolic cluster centred on diabetes and an inflammatory cluster linking respiratory and atopic conditions.}






\maketitle


\section{Introduction}
\label{sec:intro}

The increasing global burden of multiple long-term conditions (MLTCs), commonly referred to as multimorbidity, presents a major challenge to contemporary healthcare systems and public health planning \citep{barnett2012epidemiology, skou2022multimorbidity}. Multimorbidity now affects more than one-third of adults in high-income countries, and in the United Kingdom the majority of
people over 65 years of age carry at least two long-term conditions simultaneously, with prevalence increasing among younger cohorts in a manner not fully explained by demographic ageing alone \citep{barnett2012epidemiology, kuan2019chronological}. Multimorbidity is increasingly understood as a dynamic, non-random process driven by shared biological mechanisms, systemic inflammation, metabolic dysregulation, and immune-mediated pathways that unfold continuously over the life course \citep{cheng2024chronic}. A key insight motivating the present work is that the \emph{sequence} of disease acquisitions matters: the onset of one condition may substantially alter the
rate of onset of others, generating structured cascades of morbidity accumulation that are invisible to cross-sectional analyses.

Large-scale longitudinal electronic health records (EHRs), such as the linked primary-care records of the UK Biobank, provide fine-grained temporal information on the onset of the disease in hundreds of thousands of individuals \citep{hemingway2018big, kuan2019chronological}. Exploiting such data, however, requires statistical methodology capable of modelling multiple interdependent disease processes in continuous time, while accommodating irregular observation schedules, high-dimensional interaction structures, and patient-level covariate effects \citep{jackson2022comparison}. Existing approaches have important limitations: cross-sectional cluster analyses capture co-occurrence but not temporal dynamics \citep{nicholson2019multimorbidity}; multi-state and hidden Markov models encode discrete disease stages but scale poorly to large numbers of interacting conditions and do not naturally accommodate multiway interactions \citep{Putter2007}; and network-based comorbidity approaches offer intuitive visualisations but typically lack a formal probabilistic model for estimation
and uncertainty quantification \citep{valabhji2024prevalence}.

Continuous-Time Bayesian Networks (CTBNs), introduced by \citet{Nodelman2002, Nodelman2003} and formalised by \citet{nodelman2007continuous}, provide a principled probabilistic framework for these challenges. In a CTBN each
disease is modelled as a finite-state continuous-time Markov process whose transition intensities are coupled to the current states of its \emph{parent} conditions in a directed graph; that is, the disease-specific Markov processes are conditionally dependent in the sense that the instantaneous onset rate of
one condition is a function of which other conditions are currently present. This formulation accommodates irregularly spaced observations without arbitrary time discretisation, which makes CTBNs well suited to EHR-recorded disease onset
\citep{jackson2022comparison}. \citet{faruqui2021functional} extended the framework to a functional CTBN in which logarithmic transition intensities depend on patient-level covariates, with sparsity enforced through frequentist $L_1$ regularisation, and \citet{guillamet2025ctbn} developed a Bayesian CTBN
with Cox-type transition intensities for disease-progression graphs. Despite this progress, applying CTBNs to high-dimensional multimorbidity data raises two outstanding difficulties: the combinatorial explosion of parameters arising from higher-order disease interactions and the need for biologically
interpretable sparsity with calibrated uncertainty.

We address both difficulties through structured Bayesian shrinkage. The Bayesian literature offers a rich class of sparsity-inducing priors, including the Bayesian LASSO \citep{ParkCasella2008}, the horseshoe and its regularised
variant \citep{carvalho2010horseshoe, piironen2017sparsity}, and the spike-and-slab \citep{MitchellBeauchamp1988, GeorgeMcCulloch1993, rovckova2018spike}, the last of which yields posterior inclusion probabilities (PIPs) that support direct inference on the sparsity pattern. Comparative work has shown that explicit spike components tend to outperform continuous
shrinkage for hard variable selection under moderate sparsity, while continuous priors can be preferable for pure estimation
\citep{bhadra2019lasso}. Standard shrinkage priors, however, penalise all coefficients uniformly, which is sub-optimal when coefficients carry a known structure. In factorial and interaction models the \emph{hierarchical principle} \citep{HamadaWu1992} holds that higher-order interactions should be penalised more strongly than lower-order terms; this idea has been formalised through hierarchical and linked shrinkage constructions \citep{griffin2017hierarchical, van2024linked, griffin2024structured}. We embed this principle directly within the CTBN likelihood.

In this paper, we develop a structured Bayesian CTBN framework for
multimorbidity modelling in large-scale EHRs. The methodological contribution is fourfold. First, we embed four order-dependent shrinkage priors, a hierarchical structured prior, the Bayesian LASSO, the regularised horseshoe, and a continuous spike-and-slab within a unified CTBN likelihood, each applying stronger regularisation to higher-order interaction terms than to
main effects, so encoding the biological expectation that complex
multi-condition synergies are rare. Second, we place all four priors on a common selection footing by deriving PIPs for the spike-and-slab and \emph{pseudo-PIPs}, based on the posterior shrinkage factor, for the other priors. Third, we evaluated the four priors through a calibrated simulation study, isolating their parameter-recovery, variable-selection, and predictive behaviour. Fourth, we demonstrate the framework on UK Biobank primary-care records. Although the methodology is
general and applies to any set of chronic conditions, we illustrate it using a clinically coherent set of long-term inflammatory and metabolic conditions, for which we recover sparse, interpretable networks of disease progression
together with covariate effects.

The remainder of the paper is organised as follows.
Section~\ref{sec:data} describes the motivating application and study population. Section~\ref{sec:methods} introduces the CTBN model, structured shrinkage priors, posterior computation, and the predictive evaluation framework. Section~\ref{sec:sim} presents the calibrated simulation study that establishes the methodological properties of the four priors. The framework is
then applied to the UK Biobank cohort in Section~\ref{sec:empirical}, and Section~\ref{sec:discussion} discusses the findings, situates them within the related literature and concludes.

\section{Motivating Example}
\label{sec:data}

To motivate this work, we analyse longitudinal electronic health record data from the InflAIM programme, an NIHR-funded study of inflammation-associated multimorbidity. Multiple long-term conditions (MLTCs) arise through complex and evolving interactions among diseases, and recent studies have demonstrated the value of disease networks and temporal disease trajectories for characterising patterns of disease progression \citep{jensen2014temporal,siggaard2020disease}. However, existing approaches often focus on disease co-occurrence or trajectory discovery and do not jointly model disease transitions, disease-dependency networks, and high-dimensional risk factors within a unified framework.

InflAIM pre-specifies a panel of 60 inflammatory, autoimmune, and metabolic long-term conditions (the \emph{InflAIM set}; Supplementary Table~S1). The code lists used to identify the conditions are published on \href{https://github.com/Inflaim-Study/Inflaim-LTC-code-list}{GitHub}. The complete condition set was used to define the eligibility of the cohort using a disease-free baseline criterion. Participants were required to be free of all InflAIM conditions at age 40 and were subsequently followed for incident disease diagnoses. For the trajectory modelling analyses, we restricted our focus to the ten most prevalent conditions among eligible participants which include Malignant neoplasms (MN), Ischaemic heart diseases (IHD), Dermatitis and eczema (DE), Hypertensive diseases (HD), Diseases of veins and lymphatic vessels (DVL), Rhinitis and sinusitis (RS), Conductive and sensorineural hearing loss (HL), Diabetes mellitus (DM), Osteoarthritis (OA), and Chronic lower respiratory diseases (CLRD) to ensure sufficient event frequencies for reliable estimation of disease transition intensities and interactions. The resulting study population comprised 33,558 UK Biobank participants who developed at least one of the selected conditions during follow-up.

The InflAIM cohort is used throughout this paper to illustrate the proposed CTBN framework and to evaluate its ability to identify directed disease dependencies and risk factors associated with disease progression.

\section{Methods}
\label{sec:methods}

\subsection{Continuous-Time Bayesian Networks for Inflammatory Disease Trajectories}

We model the joint progression of $M$ inflammatory conditions over continuous time using a Continuous-Time Bayesian Network (CTBN). Let $\mathbf{X}(t) = (X_1(t), \dots, X_M(t))$ denote a multivariate stochastic process, where each $X_m(t)$ represents the state of condition $m$ at time $t$. Each component evolves as a finite-state continuous-time Markov process, with instantaneous transition rates that may depend on the current states of other variables. Specifically, if $X_m(t^-) = i$ and all other variables are in configuration $u$ at time $t^-$, the conditional intensity of transitioning to state $j \neq i$ is $q_{i \to j}^{(m)}(u) \geq 0$, with the holding time in state $i$ being exponentially distributed with rate $q_i^{(m)}(u) = \sum_{j \neq i} q_{i \to j}^{(m)}(u)$. For a binary variable, these rates are collected into a $2 \times 2$ conditional intensity matrix
\[
Q_{m|u} =
\begin{bmatrix}
-q_{0\to1}^{(m)}(u) & q_{0\to1}^{(m)}(u) \\
q_{1\to0}^{(m)}(u) & -q_{1\to0}^{(m)}(u)
\end{bmatrix},
\]
where rows sum to zero by construction. Since each $X_m(t) \in \{0,1\}$, the indices $i, j \in \{0,1\}$ and there are at most two distinct transitions per variable. We further restrict attention to chronic inflammatory conditions, for which remission is rare on the timescales of interest. We therefore set $q_{1 \to 0}^{(m)}(u) = 0$ for all $m$ and $u$, making the state $X_m = 1$ absorbing. The conditional intensity matrix simplifies accordingly to
\[
Q_{m|u} =
\begin{bmatrix}
-q_{0\to1}^{(m)}(u) & q_{0\to1}^{(m)}(u) \\
0 & 0
\end{bmatrix}.
\]
The only non-trivial transition is therefore onset, $0 \to 1$, and each condition is characterised by a single rate, which we denote $q_m(u)$, the conditional intensity of onset given the current configuration $u$ of all other variables. 

Having reduced the model to a single rate per condition, we turn to specifying how $q_m$ depends on the current state of the system. In full generality, the onset rate could depend on all $2^{M-1}$ binary configurations of the remaining $M-1$ conditions, as well as patient-level covariates. This is not estimable without further modelling assumptions, and we therefore adopt two complementary restrictions.

First, rather than allowing $q_m$ to depend on all other conditions, we restrict dependence to a subset $\mathrm{Pa}(m) \subseteq \{1, \dots, M\} \setminus \{m\}$, which we call the \emph{parent set} of condition $m$. Denoting by $u \in \{0,1\}^{|\mathrm{Pa}(m)|}$ the configuration of parent conditions at $t^-$, the onset rate is written $q_m(t, u)$. This sparsity assumption is encoded in a directed graph $\mathcal{G}$ over the $M$ conditions, where a directed edge $\ell \to m$ indicates that $X_\ell$ belongs to $\mathrm{Pa}(m)$ and may directly influence the onset rate of condition $m$. Restricting to parent configurations reduces the $2^{M-1}$ possible configurations to $2^{|\mathrm{Pa}(m)|}$, yielding a tractable parameterisation.

Second, we adopt a log-linear form for the onset rate, which avoids the need to enumerate all parent configurations explicitly. Letting $X_\ell(t^-)$ for $\ell \in \mathrm{Pa}(m)$ denote the parent states at $t^-$, and $\mathbf{Z}(t)$ the vector of patient covariates (which may include both time-invariant and time-varying components), we write
\begin{equation}\label{eqn1}
\log q_m(t, u) = \beta^0_m + \sum_{\ell \in \mathrm{Pa}(m)} \beta_{m,\{\ell\}}^{(1)}\, X_\ell(t^-) + \sum_{\substack{L \subseteq \mathrm{Pa}(m) \\ 2 \leq |L| \leq P}} \beta_{m,L}^{(|L|)} \prod_{\ell \in L} X_\ell(t^-) + \mathbf{Z}(t)^\top \boldsymbol{\gamma}_m,
\end{equation}
where $\beta^0_m \in \mathbb{R}$ is the baseline log-intensity, $\beta_{m,L}^{(p)} \in \mathbb{R}$ is the coefficient for the $p$-way interaction among the conditions in subset $L \subseteq \mathrm{Pa}(m)$, and $\boldsymbol{\gamma}_m$ captures covariate effects. The $p=1$ terms capture the main effects of each parent condition, while $p \geq 2$ terms capture synergistic interactions among subsets of parent conditions of size $p$. This log-linear structure ensures positivity of rates: $\exp(\beta_{m,\{\ell\}}^{(1)}) > 1$ indicates that parent condition $\ell$ accelerates onset of condition $m$, while a positive $\beta_{m,\{\ell,k\}}^{(2)}$ indicates that conditions $\ell$ and $k$ jointly accelerate onset beyond their individual effects, capturing potential biological synergy. To preserve parsimony, interaction terms are restricted to low order $P$.

The graph $\mathcal{G}$ is not assumed known in advance. Instead, we take a data-driven approach, placing shrinkage priors on the full collection of coefficients $\{\beta_{m,L}^{(p)}\}$ for all $\ell \in \{1,\dots,M\}\setminus\{m\}$ and $p \geq 1$. Shrinkage on the main effects $\beta_{m,\{\ell\}}^{(1)}$ drives edge selection in $\mathcal{G}$, while shrinkage on higher-order terms encourages parsimony in the interaction structure. This allows both the dependency graph and the interaction terms to be learned from the observed disease trajectories, as described in Section~\ref{sec:priors}.

\subsection{Likelihood Derivation and Connection to the Poisson Model}

We now derive the likelihood for the CTBN model and establish its connection to the Poisson log-likelihood over at-risk intervals, which forms the computational foundation for estimation.

\subsubsection{Data structure} Let there be $N$ patients indexed $i = 1, \dots, N$. For each patient, the observed follow-up period is partitioned into a sequence of contiguous at-risk intervals. Interval $s$ for patient $i$ spans $[t_{i,s},\, t_{i,s+1})$ with exposure width
\[
T_{i,s} = t_{i,s+1} - t_{i,s} > 0.
\]
Within each interval, the configuration of all conditions and covariates is assumed constant at the values observed at $t_{i,s}^-$. We denote by $\mathbf{X}_{i,s} \in \{0,1\}^M$ the vector of condition states and by $\mathbf{Z}_{i,s}$ the vector of covariate values for patient $i$ in interval $s$. For a target condition $m$, patient $i$ is \emph{at risk} during interval $s$ if $X_m(t_{i,s}) = 0$. The binary event indicator is

\begin{equation}
N^m_{i,s} = \mathbf{1}\{X_m(t_{i,s+1}) = 1 \mid X_m(t_{i,s}) = 0\} \in \{0, 1\}.
\end{equation}

\subsubsection{Likelihood from first principles} Under the absorbing-state CTBN of Section 3.1, the only non-trivial transition for condition $m$ is onset ($0 \to 1$), occurring at intensity $q_m(t, u)$ that depends on the current parent configuration $u = \mathbf{X}_{\mathrm{Pa}(m), i,s}$ and covariates $\mathbf{Z}_{i,s}$. Over a sufficiently short interval of length $T_{i,s}$, the probability of onset is approximately $q_m \cdot T_{i,s}$, and the standard Poisson process approximation \citep{CoxMiller1965} gives
\begin{equation}
N^m_{i,s} \mid q_{m,i,s},\, T_{i,s} \;\sim\; \mathrm{Poisson}\!\bigl(q_{m,i,s} \cdot T_{i,s}\bigr),
\end{equation}
where $q_{m,i,s} \equiv q_m(t_{i,s}, u_{i,s})$ denotes the onset intensity for condition $m$ evaluated at the configuration of patient $i$ in interval $s$. Formally, for an interval in which onset occurs (i.e.\ $N^m_{i,s} = 1$) the likelihood contribution is $q_{m,i,s} \exp(-q_{m,i,s}\, T_{i,s})$, and for a censored interval ($N^m_{i,s} = 0$) it is $\exp(-q_{m,i,s}\, T_{i,s})$. Both cases are subsumed by the Poisson form above.

\subsubsection{Log-linear intensity and the Poisson GLM} Substituting the log-linear parameterisation from equation~\eqref{eqn1}, the log-intensity for condition $m$ in interval $(i,s)$ is
\begin{equation}
\log q_{m,i,s} = \beta^{0}_m + \sum_{j \in \mathrm{Pa}(m)} \beta^m_{\{j\}} X_{j,i,s} + \sum_{\substack{L \subseteq \mathrm{Pa}(m) \\ 2 \leq |L| \leq P}} \beta^m_L \prod_{j \in L} X_{j,i,s} + \mathbf{Z}_{i,s}^\top \boldsymbol{\gamma}_m,
\end{equation}
where $X_{j,i,s}$ denotes the state of condition $j$ (appearing as a predictor) for patient $i$ in interval $s$, and $\beta^m_L$ is the interaction coefficient for subset $L$ in the model for condition $m$. Incorporating the exposure $T_{i,s}$ as a log-offset, the Poisson model becomes
\[
N^m_{i,s} \mid \boldsymbol{\beta}^m, \boldsymbol{\gamma}_m, T_{i,s} \;\sim\; \mathrm{Poisson}\!\Bigl(\exp\bigl(\log q_{m,i,s} + \log T_{i,s}\bigr)\Bigr).
\]

\subsubsection{Observed-data log-likelihood} By the Markov property and the conditional independence structure of the CTBN, the likelihood factorizes across conditions $m$, patients $i$, and intervals $s$. The full observed-data log-likelihood for condition $m$ over all $n_m$ at-risk intervals is
\begin{equation}\label{eqn:loglik}
\ell(\boldsymbol{\beta}^m, \boldsymbol{\gamma}_m \mid \mathcal{D}) = \sum_{i=1}^{N} \sum_{s \in \mathcal{S}^m_i} \Bigl[ N^m_{i,s} \bigl(\log q_{m,i,s} + \log T_{i,s}\bigr) - q_{m,i,s}\, T_{i,s} \Bigr],
\end{equation}
where $\mathcal{S}^m_i$ denotes the set of at-risk intervals for patient $i$ with respect to condition $m$, and the constant term $\log N^m_{i,s}!$ has been dropped as it does not depend on the parameters. Equation~\eqref{eqn:loglik} is precisely a Poisson GLM log-likelihood with a log-offset \citep{McCullaghNelder1989}, confirming that inference for each condition $m$ reduces to a penalised Poisson regression over the collection of at-risk intervals. The joint log-likelihood across all $M$ conditions is
\[
\ell(\{\boldsymbol{\beta}^m, \boldsymbol{\gamma}_m\}_{m=1}^M \mid \mathcal{D}) = \sum_{m=1}^{M} \ell(\boldsymbol{\beta}^m, \boldsymbol{\gamma}_m \mid \mathcal{D}),
\]
which factorizes completely across conditions, so that estimation for each $m$ may proceed independently given the graph $\mathcal{G}$. This factorization, combined with the Poisson GLM structure of each term, is what makes the model computationally tractable even for moderately large $M$.

\subsection{Bayesian Estimation with Structured Shrinkage Priors}
\label{sec:priors}

The Bayesian framework provides a principled approach to regularising the high-dimensional parameter space of the CTBN. For each target condition $m$, let $\boldsymbol{\theta}^m = (\boldsymbol{\beta}^m, \boldsymbol{\gamma}_m)$ denote the complete parameter vector, where $\boldsymbol{\beta}^m = \{\beta^m_L : L \subseteq \{1,\dots,M\}\setminus\{m\},\, |L| \leq P\}$ collects all main-effect and interaction coefficients and $\boldsymbol{\gamma}_m \in \mathbb{R}^K$ collects the $K$ covariate effects. The posterior distribution is
\[
p(\boldsymbol{\theta}^m \mid \mathcal{D}) \;\propto\; p(\mathcal{D} \mid \boldsymbol{\theta}^m)\, p(\boldsymbol{\theta}^m),
\]
where $p(\mathcal{D} \mid \boldsymbol{\theta}^m)$ is the Poisson log-likelihood derived in Section 3.2. Since the likelihood factorises across conditions, the posterior for each $m$ is independent given the graph $\mathcal{G}$, and we describe the prior specification for a single condition $m$, suppressing the superscript $m$ where no ambiguity arises.

Each coefficient $\beta_j$ is associated with an interaction order $p_j \in \{1, \dots, P\}$, where $p_j = 1$ for main (first-order) effects and $p_j = r$ for $r$-way interactions involving $r$ parent conditions. All four prior specifications are constructed to enforce stronger shrinkage on higher-order interaction coefficients, thereby encoding the biological expectation that complex synergistic pathways are sparse. This is achieved via an order-dependent effective variance that decays as $\exp(-\theta p_j)$, where $\theta > 0$ is a penalty parameter. Instead of placing a hyperprior on $\theta$, we fix it at a default value ($\theta = 1$) and interpret it as a tuning constant whose impact is evaluated explicitly: sensitivity of all inferential results to $\theta \in \{0.5, 1, 2\}$ is presented in Supplementary Figure~S1. Covariate coefficients $\boldsymbol{\gamma}_m$ are modelled separately under a weakly informative prior in all formulations, since covariate effects are not subject to the same order-based penalisation.

\subsubsection{Overview of the Four Priors}

We compare four order-dependent shrinkage priors. The full hierarchies, marginal densities, and closed-form full conditionals for the auxiliary parameters of each prior are given in Supplementary Section~S1; we summarise the four priors here.

\paragraph{Hierarchical structured (HS) prior.} A normal--inverse-gamma scale mixture in which the slab variance is order-specific, $\sigma^2_p \sim \mathrm{Inv\text{-}Gamma}(a_0, b_0) / \Gamma(p+1)$. Marginalising the variance yields a scaled Student-$t$ prior on $\beta_j$ with scale shrinking with $p_j$, giving smooth continuous regularisation but no hard variable selection.

\paragraph{Bayesian LASSO.} A scale mixture of normals with exponential mixing distribution, yielding a double-exponential (Laplace) marginal for $\beta_j$ with order-dependent rate \citep{ParkCasella2008}. A per-order penalty parameter $\lambda^2_p$ allows the data to inform the strength of regularisation at each interaction order.

\paragraph{Regularised horseshoe.} A global--local prior $\beta_j = z_j\,\tilde{\lambda}_j\,\tau_j$ with $z_j \sim \mathcal{N}(0,1)$, half-Cauchy local scale $\lambda_j$, and order-penalised global scale $\tau_j = \tau_0\, e^{-\theta p_j}$; the slab variance $c^2$ stabilises the tails \citep{piironen2017sparsity}. The prior gives adaptive shrinkage, a sharp spike near zero with heavy tails, and is appropriate when a small number of strong signals coexists with many near-zero coefficients.

\paragraph{Continuous spike-and-slab.} A two-component normal mixture in which each coefficient is drawn from a diffuse slab with probability $\pi_{p_j}$ or a concentrated spike with probability $1 - \pi_{p_j}$,
\[
\beta_j \;\sim\; \pi_{p_j}\,\mathcal{N}\!\left(0,\; \sigma^2_{p_j}\right) + (1 - \pi_{p_j})\,\mathcal{N}\!\left(0,\; \epsilon\right),
\qquad
\pi_{p_j} = \pi_0\, e^{-\theta p_j},
\]
with order-specific slab variance $\sigma^2_{p_j} \sim \mathrm{Inv\text{-}Gamma}(a_0, b_0)$, fixed spike variance $\epsilon$, and base inclusion probability $\pi_0 \in (0,1)$.We use the marginalised mixture above with the \emph{deterministic} order-dependent inclusion probability $\pi_{p_j}$, so the order penalty enters through the inclusion probability rather than the slab variance. The posterior inclusion probability for $\beta_j$ is recovered analytically at each MCMC draw via Bayes' theorem:
\begin{equation}\label{eq:pip}
\mathrm{PIP}_j = \frac{\pi_{p_j}\,\phi(\beta_j;\, 0, \sigma^2_{p_j})}{\pi_{p_j}\,\phi(\beta_j;\, 0, \sigma^2_{p_j}) + (1 - \pi_{p_j})\,\phi(\beta_j;\, 0, \epsilon)},
\end{equation}
where $\phi(\cdot;\, 0, v)$ is the $\mathcal{N}(0, v)$ density. A directed edge $j \to m$ is declared active when $\mathbb{E}[\mathrm{PIP}_j \mid \mathcal{D}] \geq \pi^* = 0.5$.

\subsubsection{Pseudo-PIPs and Edge Selection for the Continuous Priors}
\label{sec:pseudo_pip}

The HS, Bayesian LASSO, and regularised horseshoe priors do not yield explicit posterior inclusion probabilities (PIPs). To place all four priors on a common basis for variable selection, we therefore construct a posterior \emph{pseudo-PIP} for each coefficient $\beta_j$ under the continuous priors by mapping the posterior shrinkage factor $\kappa_j \in [0,1]$ to an inclusion score. Following \citet{carvalho2010horseshoe} and \citet{piironen2017sparsity}, $\kappa_j$ is defined as the proportion of the (Gaussian/Laplace-approximate) conditional posterior mean of $\beta_j$ that is attributable to the prior, i.e.\ it is the shrinkage weight in
\[
\mathbb{E}[\beta_j \mid \mathrm{rest}] = (1 - \kappa_j)\,\hat{\beta}_j,
\]
where $\hat{\beta}_j$ denotes the data-only (local maximum-likelihood) estimate. For the Poisson CTBN likelihood this weight assumes the familiar per-coefficient form
\begin{equation}\label{eq:kappa}
\kappa_j = \frac{1}{1 + \mathcal{I}_j\, v_j}, \qquad
\mathcal{I}_j = \sum_{i,s} x^2_{j,i,s}\, T_{i,s}\, e^{\eta_{i,s}},
\end{equation}
where $\mathcal{I}_j$ is the \emph{per-coefficient} Poisson Fisher information (i.e.\ the $j$th diagonal element of $\mathbf{X}^\top \mathrm{diag}(\boldsymbol{\mu})\mathbf{X}$, with expected counts $\mu_{i,s} = T_{i,s} e^{\eta_{i,s}}$), $\eta_{i,s} = \log q_{i,s}$ and $v_j$ denotes the effective per-coefficient variance of $\beta_j$. The effective variance is given by
\[
v_j = \begin{cases}
\sigma^2_{p_j}\, e^{-\theta p_j} & \text{(hierarchical structured),}\\[4pt]
2\,\lambda^2_{p_j}\, e^{-\theta p_j} & \text{(Bayesian LASSO),}\\[4pt]
\tilde{\lambda}^2_j\, \tau^2_j & \text{(regularised horseshoe),}
\end{cases}
\]
with the third case corresponding to the local–global variance decomposition of \citet{piironen2017sparsity}. Equation~\eqref{eq:kappa} serves as the Poisson–GLM analogue of the \citet{piironen2017sparsity} shrinkage weight, which, in the orthonormal Gaussian setting with noise variance $\sigma^2$, simplifies to $\kappa_j = (1 + (n/\sigma^2)\,v_j)^{-1}$. In this generalization, the homogeneous data precision term $n/\sigma^2$ is replaced by the exposure-weighted, predictor-specific information quantity $\mathcal{I}_j$. A complete derivation is provided in Supplementary Section~S1.6. 

A key feature of this construction is that $\mathcal{I}_j$ is coefficient-specific and, in the case of the horseshoe prior, $v_j$ is likewise coefficient-specific via $\tilde{\lambda}_j$. Consequently, $\kappa_j$ varies across coefficients even when they correspond to interactions of the same order: two parent terms of identical interaction order but differing empirical support no longer receive identical shrinkage. This reinstates the utility of $\kappa_j$ as a device for coefficient-level selection. Values of $\kappa_j$ close to zero ($\kappa_j \approx 0$) indicate that the likelihood dominates the prior, suggesting that $\beta_j$ is likely to be non-zero, whereas values $\kappa_j \approx 1$ correspond to near-complete prior-induced shrinkage. 

We define the pseudo–posterior inclusion probability (pseudo-PIP) as $\widetilde{\mathrm{PIP}}_j = \mathbb{E}[1 - \kappa_j \mid \mathcal{D}]$, which is estimated by averaging $1 - \kappa_j$ over MCMC samples. For edge selection, we employ the same decision rule $\widetilde{\mathrm{PIP}}_j \geq 0.5$ as used for the PIP under the spike-and-slab prior. We stress that the pseudo-PIP is intended as a \emph{ranking and selection device} placing the continuous-shrinkage priors on a footing comparable to the spike-and-slab, and not as a formal posterior probability of inclusion: unlike the PIP of equation~\eqref{eq:pip}, which arises from an explicit two-component mixture, $\widetilde{\mathrm{PIP}}_j$ is a monotone transformation of the posterior shrinkage weight and should be read as a relative importance score rather than a calibrated inclusion probability. With this caveat, the unified scoring framework permits the true positive rate (TPR), false positive rate (FPR), and selection AUC to be computed in a directly comparable manner across all four priors considered in the simulation study (Section~\ref{sec:sim}).

\subsubsection{Posterior Computation}
The coefficient vector $\boldsymbol{\beta}^m$ enters the Poisson likelihood non-conjugately under all four priors, so the full conditionals for $\beta_j$ are not available in closed form. Posterior sampling is performed by NUTS in Stan; auxiliary scale parameters retain closed-form Gibbs updates as detailed in Supplementary Section~S1. A summary comparison of the four priors is given in Table~\ref{tab:priors}.

\begin{table}[h]
\centering
\caption{Comparison of shrinkage priors for CTBN estimation. $\widetilde{\mathrm{PIP}}_j = \mathbb{E}[1 - \kappa_j \mid \mathcal{D}]$ denotes the pseudo-PIP derived from the posterior shrinkage factor.}
\label{tab:priors}
\begin{tabular}{lllll}
\hline
Prior & Sparsity & Hard selection & Output & Recommended when \\
\hline
Hierarchical structured & Soft & No & $\widetilde{\mathrm{PIP}}_j$, CI & Smooth shrinkage \\
Bayesian LASSO & Moderate & No & $\widetilde{\mathrm{PIP}}_j$, CI & Many small effects expected \\
Regularised horseshoe & Adaptive & No & $\widetilde{\mathrm{PIP}}_j$, CI & Few large signals \\
Spike-and-slab & Hard & Yes & PIP & Explicit network selection \\
\hline
\end{tabular}
\end{table}
The shared hyperparameters are $\theta$ (order penalty), $a_0, b_0$ (inverse-gamma shape and scale), and $\pi_0$ (base inclusion probability for spike-and-slab). Sensitivity to $\theta$ is reported in the empirical results.

\subsection{Survival and Cumulative Incidence Functions}

For patient $i$ at risk for condition $m$ at entry $t_0$, with parent condition 
states and covariates evolving along observed path $\{(\mathbf{x}_{i,s}, 
\mathbf{z}_{i,s})\}$ across intervals $s = 1, \dots, S_i$, the 
\textbf{condition-specific survival function} is
\[
S_m(t \mid \mathbf{x}_i, \mathbf{z}_i)
= \exp\!\left(-\int_{t_0}^{t} q_m(u,\, \mathbf{x}_i(u),\, \mathbf{z}_i(u))\,
\mathrm{d}u\right).
\]
Since the intensity is piecewise-constant over at-risk intervals, the integral 
reduces to
\[
S_m(t \mid \mathbf{x}_i, \mathbf{z}_i)
= \exp\!\left(-\sum_{s:\, t_{i,s} < t} q_{m,i,s}\cdot T_{i,s}^*\right),
\]
where $T_{i,s}^* = \min(t_{i,s+1}, t) - t_{i,s}$ is the time spent in interval 
$s$ up to horizon $t$, and
\[
q_{m,i,s}
= \exp\!\left(\mathbf{x}_{i,s}^\top \boldsymbol{\beta}^m
+ \mathbf{z}_{i,s}^\top \boldsymbol{\gamma}_m\right).
\]
The \textbf{cumulative incidence function} is the complement,
\[
F_m(t \mid \mathbf{x}_i, \mathbf{z}_i)
= 1 - S_m(t \mid \mathbf{x}_i, \mathbf{z}_i).
\]
Since each condition has a single absorbing onset event with no remission, this coincides with the cause-specific CDF and no subdistribution adjustment is required \citep{FineGray1999, Putter2007}. Under the Bayesian framework, the posterior mean survival and cumulative incidence functions are obtained by Monte Carlo integration over 
$T$ posterior draws $\{(\boldsymbol{\beta}^{m(t)}, \boldsymbol{\gamma}_m^{(t)}
)\}_{t=1}^T$:
\[
\bar{S}_m(t \mid \mathbf{x}_i, \mathbf{z}_i)
\approx \frac{1}{T}\sum_{t=1}^T
\exp\!\left(-\sum_{s:\, t_{i,s} < t} q^{(t)}_{m,i,s}\cdot T_{i,s}^*\right),
\qquad
\bar{F}_m = 1 - \bar{S}_m.
\]
This estimator is used 
in the predictive performance assessments of Section~3.5.

\section{Simulation Study}\label{sec:sim}

To rigorously evaluate the performance of the CTBN framework with structured shrinkage priors, we conducted a simulation study designed to mirror the empirical data structure of the UK Biobank analysis while enabling ground truth comparisons. The simulation parameters were calibrated to estimates derived from the real-data analysis using the spike-and-slab prior, ensuring ecological validity under controlled conditions.

\subsection{Simulation Design and Network Structure}

The simulation network comprises ten binary conditions corresponding to those analysed in the empirical study: Malignant neoplasms (MN), Ischaemic heart diseases (IHD), Dermatitis and eczema (DE), Hypertensive diseases (HD), Diseases of veins and lymphatic vessels (DVL), Rhinitis and Sinusitis (RS), Conductive and sensorineural hearing loss (HL), Diabetes (DM), Osteoarthritis (OA), and Chronic lower respiratory diseases (CLRD). Consistent with the notation of Section~3.1, each condition $X_m(t) \in \{0,1\}$ for $m = 1, \dots, 10$ evolves as a semi-absorbing continuous-time Markov process. Patient $i$, condition $m$, and time interval $s$ are indexed as throughout Sections~3.1--3.4.

The onset intensity for condition $m$ in interval $(i,s)$ follows the log-linear parameterisation of equation~\eqref{eqn1} with interactions up to order $P = 2$:
\[
\log q_{m,i,s} = \beta^0_m + \sum_{j \in \mathrm{Pa}(m)} \beta^m_{\{j\}}\, X_{j,i,s} + \sum_{\substack{L \subseteq \mathrm{Pa}(m) \\ |L| = 2}} \beta^m_L \prod_{j \in L} X_{j,i,s} + \mathbf{Z}_{i,s}^\top \boldsymbol{\gamma}_m,
\]
where $\beta^0_m$ is the baseline log-intensity, $\beta^m_{\{j\}}$ captures the main effect of parent condition $j$ on the onset of condition $m$, $\beta^m_L$ models the pairwise synergistic interaction between the two conditions in $L \subseteq \mathrm{Pa}(m)$, and $\boldsymbol{\gamma}_m$ encodes covariate effects. The absorbing-state intensity matrix for each condition is $Q_{m|u} = \bigl(\begin{smallmatrix} -q_{m,i,s} & q_{m,i,s} \\ 0 & 0 \end{smallmatrix}\bigr)$, reflecting the chronic nature of the conditions as established in Section~3.1.

\subsubsection{Parameter Calibration from Empirical Estimates}

Parameter values were calibrated to align with empirical estimates from the spike-and-slab model. Baseline log-intensities $\beta^0_m$ were set to approximate observed one-year incidence rates from the UK Biobank, ranging from $\exp(\beta^0_m) = 0.005$ per year for Malignant neoplasms to $\exp(\beta^0_m) = 0.15$ per year for Hypertensive diseases. The interaction structure was sparse: approximately 60\% of possible pairwise coefficients $\beta^m_L$ were set to zero. The remaining 40\% received non-zero values drawn from a mixture: with probability 0.7, $\beta^m_L \sim \mathcal{N}(0, 0.3^2)$, representing weak to moderate synergistic effects, and with probability 0.3, $\beta^m_L \sim \mathcal{N}(1.2, 0.1^2)$, capturing stronger synergistic relationships consistent with known clinical comorbidity patterns.

The covariate vector $\mathbf{Z}_{i,s}$ for patient $i$ in interval $s$ contains three predictors. Age was modelled as a continuous time-varying covariate with initial distribution $\mathrm{age}_{i,0} \sim \mathcal{N}(57, 9.39^2)$ and deterministic progression $\mathrm{age}_{i,s} = \mathrm{age}_{i,0} + t_{i,s}$, where $t_{i,s}$ is the calendar time at the start of interval $s$. Sex was treated as a binary time-invariant covariate with $\Pr(\mathrm{male}) = 0.56$. Smoking status was treated as a time-varying categorical variable with three ordered states: never (0), former (1), and current (2) smoker. To respect the irreversibility of smoking cessation and the impossibility of reverting to never-smoker status, transitions are restricted to the one-directional process current $\to$ former and never $\to$ current, giving the intensity matrix
\[
Q_{\mathrm{smoking}} =
\begin{pmatrix}
-(q_{01}^{\mathrm{s}}) & q_{01}^{\mathrm{s}} & 0 \\
0 & -(q_{12}^{\mathrm{s}}) & q_{12}^{\mathrm{s}} \\
0 & 0 & 0
\end{pmatrix},
\]
where rows and columns correspond to never, current, and former smoking states respectively, $q_{01}^{\mathrm{s}} = 0.04$ is the rate of smoking initiation (never $\to$ current) and $q_{12}^{\mathrm{s}} = 0.06$ is the rate of cessation (current $\to$ former); only these two transitions are permitted, since never $\to$ former is biologically implausible and the former state is absorbing, reflecting the permanent nature of having quit. The initial smoking distribution was set to $\mathrm{Categorical}(0.47, 0.11, 0.42)$ for never, current, and former respectively, calibrated to UK Biobank baseline prevalences. Covariate effects $\boldsymbol{\gamma}_m$ are reported in Supplementary Table~S6, with age effects positive across all conditions and strongest for cardiovascular and metabolic diseases, and smoking effects most pronounced for respiratory and malignant conditions.

\subsubsection{Data Generation Mechanism}

We generated $N = 5{,}000$ independent patient trajectories over the time horizon $[0, 5]$ years. Initial condition states were drawn from Bernoulli distributions with prevalences calibrated to UK Biobank estimates: MN (0.06), IHD (0.08), DE (0.07), HTN (0.16), DVL (0.04), RS (0.08), HL (0.04), DM (0.05), OA (0.06), and CLRD (0.05). Initial covariate values were drawn from the distributions described in Section~4.1.1.

Patient trajectories were simulated using the following exact procedure, which exploits the piecewise-constant structure of the CTBN intensity. At any time $t$, the joint onset intensity across all conditions still in state 0 is
\[
\Lambda_i(t) = \sum_{m:\, X_{m,i}(t) = 0} q_{m,i}(t),
\]
where $q_{m,i}(t) = \exp(\log q_{m,i,s})$ for the interval $s$ containing $t$. Given the current system state, the waiting time to the next event is exponentially distributed, $W_i \sim \mathrm{Exp}(\Lambda_i(t))$, the exact distribution of the next event time in a finite-state continuous-time Markov chain by the memoryless property \citep{Norris1997}. Because covariate updates and condition onsets change the intensity between intervals, we draw events by \emph{thinning} \citep{Lewis1979, Ogata1981}: a candidate time is proposed at a dominating rate $\bar{\Lambda} \geq \Lambda_i(t)$ and accepted with probability $\Lambda_i(t^*)/\bar{\Lambda}$. On acceptance at $t^*$, the transitioning condition is drawn with probability $q_{m,i}(t^*)/\Lambda_i(t^*)$, the multinomial selection step of the Gillespie algorithm \citep{Gillespie1977}, after which parent configurations and intensities of the affected child conditions are updated. The recorded event histories are converted to the at-risk interval format of Section~3.2 for model fitting.

\subsection{Performance Evaluation Metrics}

Model performance is assessed under two complementary paradigms reflecting distinct inferential objectives: (i) parameter recovery and variable selection, and (ii) predictive accuracy. Parameter recovery and variable selection are evaluated using a Monte Carlo design with $R = 100$ independent replicates. In each replicate, a dataset of $N = 5{,}000$ patients is generated from the data-generating mechanism in Sections~4.1--4.3, and the model is fitted to the full dataset. This design yields stable estimates of bias, RMSE, and coverage that are not driven by a single realisation, consistent with standard practice in Bayesian simulation studies \citep{morris2019bayesian}. For each replicate, posterior summaries are obtained from NUTS samples and performance metrics are averaged across replicates. Predictive performance is evaluated using $K = 5$-fold cross-validation within each replicate. Patients are partitioned at the individual level to prevent information leakage across time intervals. Models are trained on $K-1$ folds and evaluated on held-out data, with fold-level metrics averaged within replicates and then across replicates.

\subsubsection{Parameter recovery}

Parameter recovery is assessed on the full simulated dataset within each replicate. For each parameter $\omega = \{{\beta^0_m, \beta^m_{{j}}, \beta^m_L, \gamma_{m,k}}\}$, we report bias, root mean squared error (RMSE), and 95\% credible interval coverage:
\[
\mathrm{Bias}(\hat{\omega}) = \hat{\omega} - \omega^{\mathrm{true}}, \qquad
\mathrm{RMSE}(\hat{\omega}) = \sqrt{\left(\hat{\omega} - \omega^{\mathrm{true}}\right)^2 + \widehat{\mathrm{Var}}(\hat{\omega})},
\]
\[
\mathrm{Coverage}_{95\%}(\hat{\omega}) = \mathbf{1}\!\left(\omega^{\mathrm{true}} \in \mathrm{CI}_{0.95}(\hat{\omega})\right),
\]
Here, $\hat{\omega}$ denotes the posterior mean and $\widehat{\mathrm{Var}}(\hat{\omega})$ the posterior variance obtained from NUTS samples. Metrics are computed separately for baseline, main effects, interactions, and covariate effects, and then summarised across parameters within each class and condition.

\subsubsection{Variable selection}

Variable selection performance is evaluated using the full-data posterior in each replicate. For the spike-and-slab prior, posterior inclusion probabilities (PIPs) are computed from equation~\eqref{eq:pip}. For continuous shrinkage priors (horseshoe, regularised horseshoe, Bayesian LASSO), pseudo-PIPs defined in Section~\ref{sec:pseudo_pip} are used. A common threshold of 0.5 is applied across all priors to declare edge inclusion, enabling fair comparison. Let $\mathcal{E}$ denote the set of candidate edges or coefficients. Within replicate $r$, true and false positive rates are defined as
\[
\mathrm{TPR}^{(r)} = \frac{\sum_{j \in \mathcal{E}} \mathbf{1}\{\widehat{\mathrm{PIP}}_j^{(r)} > 0.5\}\, \mathbf{1}\{\beta_j^{\mathrm{true}} \neq 0\}}{\sum_{j \in \mathcal{E}} \mathbf{1}\{\beta_j^{\mathrm{true}} \neq 0\}},
\qquad
\mathrm{FPR}^{(r)} = \frac{\sum_{j \in \mathcal{E}} \mathbf{1}\{\widehat{\mathrm{PIP}}_j^{(r)} > 0.5\}\, \mathbf{1}\{\beta_j^{\mathrm{true}} = 0\}}{\sum_{j \in \mathcal{E}} \mathbf{1}\{\beta_j^{\mathrm{true}} = 0\}}.
\]
Selection AUC is obtained by varying the inclusion threshold over PIPs. Metrics are reported separately for main effects ($|L|=1$) and interactions ($|L|=2$), and averaged across replicates.

\subsubsection{Predictive accuracy}

Predictive accuracy is evaluated using held-out test folds within each replicate. For each condition, we report the Poisson log-likelihood and interval Brier score, alongside the incident/dynamic time-dependent AUC described in \cite{Blanche2013pkg,BlancheCommengesJacqmin-Gadda2013}. For each metric, results are averaged over folds and replicates, with standard errors computed across folds. To provide a reference for calibration, we additionally compute an oracle benchmark using the true data-generating parameters on the same test folds. The ratio of the estimated-model Brier score to the oracle Brier score provides a normalised efficiency measure that is comparable across conditions with different baseline event rates.

\subsection{Analysis Models for Robustness Assessment}
\label{sec:analysis_models}

To evaluate robustness, we fit three \emph{analysis models}, each using the same Poisson CTBN likelihood but differing in the maximum interaction order allowed in the linear predictor:
\begin{itemize}[leftmargin=*,nosep]
\item \textbf{Analysis Model~A1 (under-specified likelihood):} $P_{\mathrm{fit}} = 1$ (main effects only), while the data-generating mechanism contains two-way interactions ($P_{\mathrm{true}} = 2$).
\item \textbf{Analysis Model~A2 (correctly specified likelihood):} $P_{\mathrm{fit}} = P_{\mathrm{true}} = 2$.
\item \textbf{Analysis Model~A3 (over-specified likelihood):} $P_{\mathrm{fit}} = 3$, including spurious three-way terms beyond the truth ($P_{\mathrm{true}} = 2$).
\end{itemize}
Each analysis model is fitted under each of the four shrinkage priors of Section~3.3, giving a $3 \times 4$ factorial of likelihood-by-prior combinations. The same Monte Carlo replicates and cross-validation folds are used across all combinations to ensure paired comparisons.

\subsection{Simulation Results}

\subsubsection{Parameter recovery results}
\begin{figure}[ht]
\centering
\includegraphics[width=\linewidth]{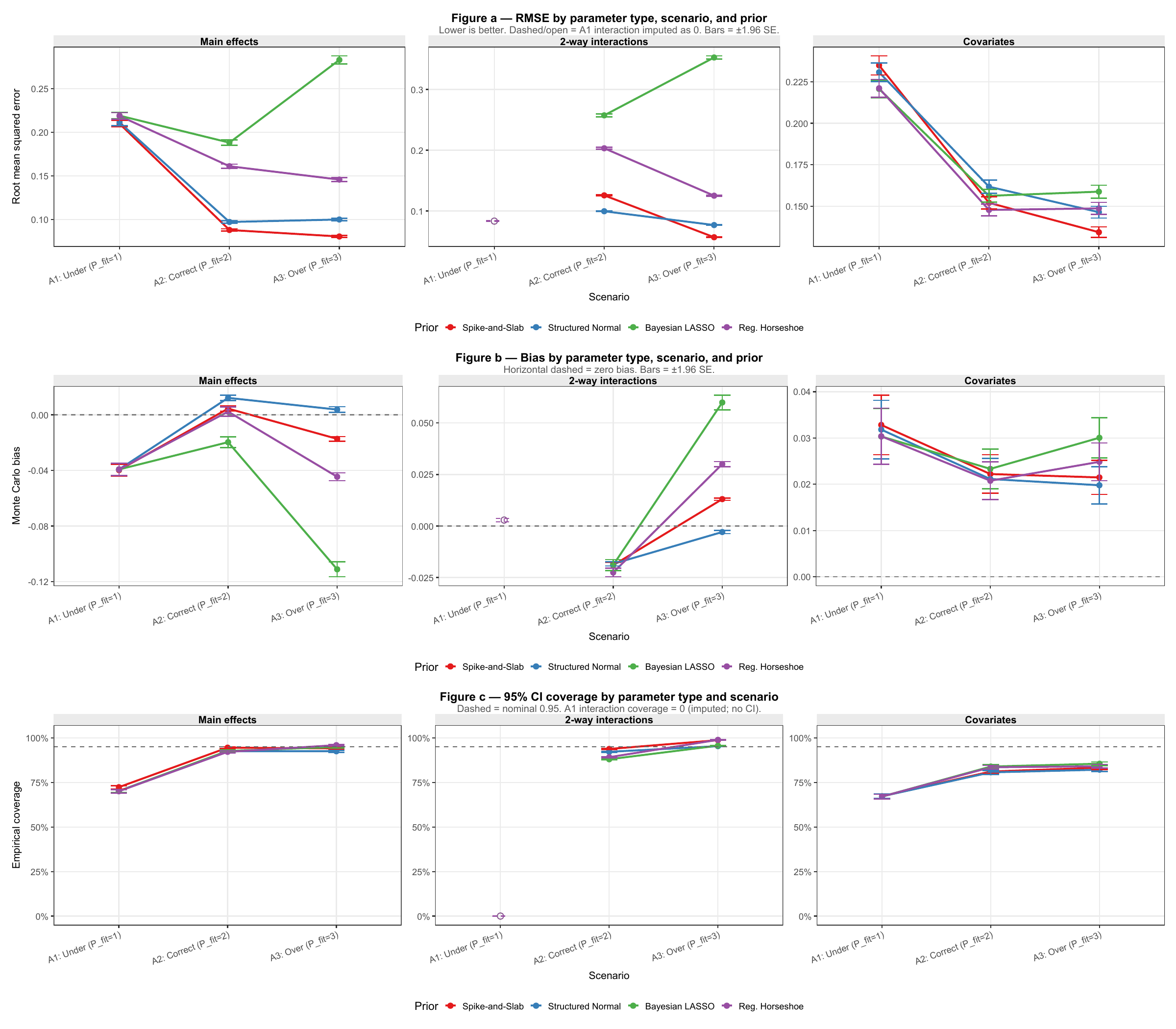}
\caption{Coefficient recovery by type, analysis model, and prior.  \emph{Top}: RMSE
(lower is better).  \emph{Middle}: Monte Carlo bias; dashed line at zero.
\emph{Bottom}: empirical coverage of 95\% credible intervals; dashed line at
nominal 0.95 level.  Columns: main-effect coefficients ($\beta$), two-way interaction coefficients,
covariate coefficients ($\gamma$). A1, A2, A3 refer to the under-, correctly,
and over-specified analysis models of Section~\ref{sec:analysis_models}.}
\label{fig:bias_coverage}
\end{figure}

Figure~\ref{fig:bias_coverage} profiles RMSE, Monte Carlo bias, and empirical 95\% credible-interval coverage across the three analysis models (A1: under-specified, A2: correctly specified, A3: over-specified), the four prior families, and the three coefficient types (main-effect, two-way interaction, and covariate coefficients). Under the under-specified model A1, main-effect RMSE is visibly elevated for all priors and Monte Carlo bias is negative, reflecting classical omitted-variable bias from the unmodelled interactions; this inflation is largest for the Bayesian LASSO and regularised horseshoe and smallest for the spike-and-slab and structured normal. Because interaction terms are not estimated under A1, their RMSE equals $|\beta^{\mathrm{true}}|$ and their coverage is zero by construction (dashed open and zero-coverage markers in Figure~\ref{fig:bias_coverage}).

Under the correctly specified model A2, all four priors yield low RMSE for every coefficient type, with the spike-and-slab attaining the smallest main-effect and interaction errors and the Bayesian LASSO the largest, though differences are modest; bias is negligible and coverage is close to (slightly below) the nominal $0.95$ for most priors, the structured normal being most consistent and the spike-and-slab modestly conservative owing to its bimodal posterior. Under the over-specified model A3, the spike-and-slab and structured normal retain near-A2 RMSE through adaptive shrinkage of the spurious higher-order terms, whereas the Bayesian LASSO and regularised horseshoe show moderate RMSE inflation for interactions; bias remains near zero throughout, so over-specification introduces variance rather than systematic bias, and interaction-coefficient coverage is most stable for the spike-and-slab.

\subsubsection{Variable selection results}
\label{sec:sim_selection}
Variable selection is summarised using a uniform $0.5$ threshold on the PIP (spike-and-slab) or on the pseudo-PIP $\widetilde{\mathrm{PIP}}_j = \mathbb{E}[1 - \kappa_j \mid \mathcal{D}]$ (structured normal, Bayesian LASSO, regularised horseshoe), with TPR, FPR, and threshold-free selection AUC reported separately for main-effect and two-way interaction coefficients across the three analysis models (Figure~\ref{fig:selection}; AUC in Supplementary Figure~S9).

The three continuous shrinkage priors cannot perform hard selection at this threshold: their posterior shrinkage factors rarely cross $0.5$, so essentially all coefficients are declared active (TPR $\approx$ FPR $\approx 100\%$) for both main effects and interactions, demonstrating that priors lacking an explicit discrete selection mechanism do not support hard variable selection at a common cutoff. The spike-and-slab is the only specification that separates signal from noise. For main effects it attains TPR $\approx 76\text{--}82\%$ with FPR falling from $\approx 13\%$ (A1) to $\approx 4\%$ (A2) and $\approx 1\%$ (A3); for two-way interactions it achieves TPR $\approx 72\%$ at FPR $\approx 4\%$ under A2, with TPR declining to $\approx 36\%$ under A3 while FPR remains near zero, indicating that selectivity is preserved under over-specification (interaction TPR is zero for all priors under A1 by construction).

\begin{figure}[H]
\centering
\includegraphics[width=\linewidth]{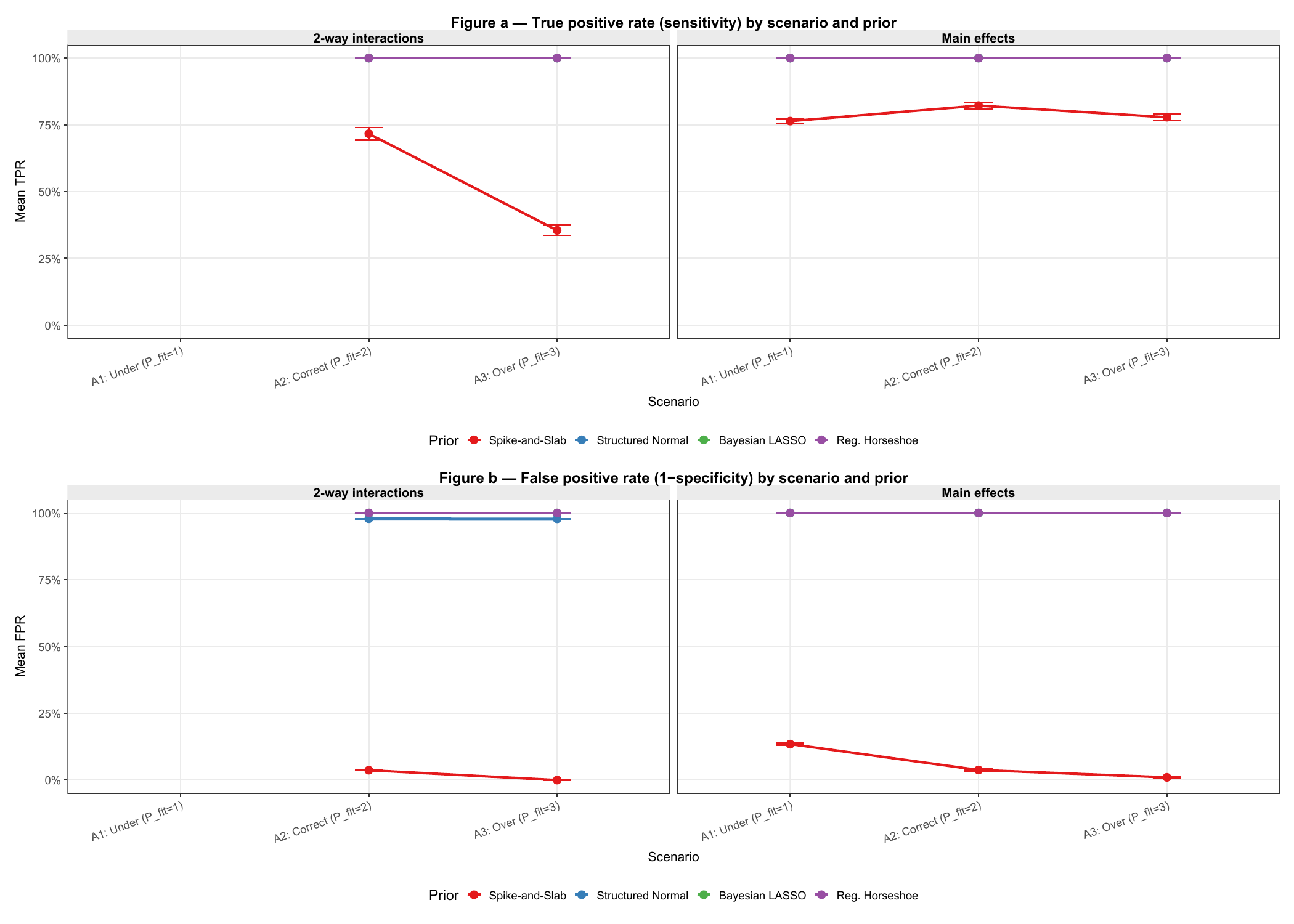}
\caption{Variable selection performance by analysis model and prior, using a uniform $0.5$ threshold on the (pseudo-)PIP for all four priors.  \emph{Top row}:
TPR.  \emph{Bottom row}: FPR.  Left column: main-effect coefficients; right column: two-way interaction coefficients.
Error bars: $\pm1.96$\,SE across Monte Carlo replicates.}
\label{fig:selection}
\end{figure}

Threshold-free selection AUC (Supplementary Figure~S9) confirms this ordering. All four priors exceed the random benchmark of $50\%$, so the (pseudo-)PIP scores carry non-trivial information even where the $0.5$ threshold fails to exploit it, but the spike-and-slab dominates: it attains AUC $\approx 85\%$ for main effects under A1, rising to $\approx 94\text{--}95\%$ under A2--A3, and $\approx 90\text{--}91\%$ for two-way interactions, whereas the three continuous priors are mutually indistinguishable at $\approx 68\text{--}71\%$. Thus only the spike-and-slab combines effective hard selection with the strongest PIP-based ranking; the continuous priors retain informative but modest rankings that cannot be converted into sparse selection at a common threshold.

\subsubsection{Predictive performance and prior comparison summary}

Table~\ref{tab:pred_metrics} reports the Poisson log-likelihood and Brier score for each prior under each analysis model. The four priors are nearly indistinguishable on held-out prediction within each analysis model: under A2, Poisson log-likelihood ranges only from $-0.5726$ (spike-and-slab) to $-0.5731$ (Bayesian LASSO) and Brier scores span $0.1573$--$0.1576$. The under-specified analysis model A1 produces the largest degradation (Poisson LL $\approx -0.5746$, Brier $\approx 0.158$ for all priors), while the over-specified A3 is marginally better than A2 for most priors except the Bayesian LASSO, which shows mild deterioration consistent with its tendency to over-fit under an expanded parameter space. Oracle efficiency analysis (Brier ratio fitted to the oracle model; Supplementary Figure~S8) confirms that all four priors achieve oracle-level efficiency across all conditions and analysis models, with no systematic left-tail mass below $0.9$.

\begin{table}[H]
\centering
\caption{Predictive performance metrics across analysis models and prior families. Poisson log-likelihood (Pois.\ LL) and Brier score are averaged over simulation replicates; standard errors (SE) are shown in parentheses. Higher Poisson log-likelihood (less negative) and lower Brier score indicate better 
predictive performance. A1: under-specified ($P_{\mathrm{fit}}=1$, interaction terms omitted); A2: correctly specified ($P_{\mathrm{fit}}=2$); A3: over-specified ($P_{\mathrm{fit}}=3$, spurious higher-order terms included).} \label{tab:pred_metrics}
\begin{tabular}{clcc}
\toprule
\multicolumn{1}{c}{Scenario} & \multicolumn{1}{c}{Prior} 
    & Pois.\ LL & Brier score \\
 & & \multicolumn{1}{c}{mean (SE)} & \multicolumn{1}{c}{mean (SE)} \\
\midrule
\multirow{4}{*}{A1: Under}
  & Spike-and-Slab       & $-0.5746$ $(0.0007)$ & $0.1582$ $(0.0002)$ \\
  & Structured Normal    & $-0.5746$ $(0.0007)$ & $0.1582$ $(0.0002)$ \\
  & Bayesian LASSO       & $-0.5746$ $(0.0007)$ & $0.1581$ $(0.0002)$ \\
  & Reg.\ Horseshoe      & $-0.5746$ $(0.0007)$ & $0.1581$ $(0.0002)$ \\
\midrule
\multirow{4}{*}{A2: Correct}
  & \textbf{Spike-and-Slab}    & $\mathbf{-0.5726}$ $(0.0007)$ & $0.1574$ $(0.0002)$ \\
  & Structured Normal          & $-0.5729$ $(0.0007)$ & $0.1576$ $(0.0002)$ \\
  & Bayesian LASSO             & $-0.5731$ $(0.0007)$ & $\mathbf{0.1573}$ $(0.0002)$ \\
  & Reg.\ Horseshoe            & $-0.5729$ $(0.0007)$ & $0.1574$ $(0.0002)$ \\
\midrule
\multirow{4}{*}{A3: Over}
  & \textbf{Spike-and-Slab}    & $\mathbf{-0.5721}$ $(0.0007)$ & $\mathbf{0.1572}$ $(0.0002)$ \\
  & Structured Normal          & $-0.5725$ $(0.0007)$ & $0.1574$ $(0.0002)$ \\
  & Bayesian LASSO             & $-0.5744$ $(0.0008)$ & $0.1574$ $(0.0002)$ \\
  & Reg.\ Horseshoe            & $-0.5725$ $(0.0007)$ & $\mathbf{0.1572}$ $(0.0002)$ \\
\bottomrule
\multicolumn{4}{l}{}
\end{tabular}

\end{table}

Combining parameter recovery, variable selection, and predictive metrics, the spike-and-slab presents the most coherent profile for network structure learning: it ranks first on RMSE, FPR, coverage calibration, and Poisson log-likelihood across all or most analysis models, at the cost of reduced TPR (rank 4) reflecting the conservatism of its hard-thresholding selection mechanism. The structured normal is a viable secondary choice when a continuous prior is preferred and hard selection is not required. The Bayesian LASSO ranks last in RMSE and FPR but achieves competitive predictive performance under A1 and A2; the structured normal, Bayesian LASSO and regularised horseshoe are all weak for variable selection (selection AUC $\approx 68\text{--}71\%$), so, despite their competitive coverage and Brier scores, none is recommended where recovery of network structure is the goal. A full rank-by-metric summary across analysis models is provided as a prior-ranking heat-map in Supplementary Figure~S13, with detailed per-prior commentary in Supplementary Section~S3.

\section{Application}
\label{sec:empirical}

We apply the proposed CTBN framework to linked primary-care electronic health records (EHRs) from the UK Biobank, a prospective cohort of approximately 500,000 UK adults recruited between 2006 and 2010. For participants who consented to general-practice (GP) record linkage, each record contains a participant identifier, event date, and coded diagnosis, and is linked to baseline demographic, socioeconomic, lifestyle, and biomarker data, including inflammatory and immune markers (C-reactive protein; white blood cell, platelet, neutrophil, and lymphocyte counts), metabolic markers (glycated haemoglobin, glycoprotein acetylation [GlycA], and albumin), and sociodemographic characteristics (sex, ethnicity, immigration status, education, employment, housing tenure, Townsend deprivation index, Dietary Inflammatory Index, body mass index category, and smoking status).

\subsection{Application methods}

The target population comprised individuals in the UK Biobank InflAIM cohort described in Section~\ref{sec:data} who were free of all 60 InflAIM conditions at age 40 and who subsequently developed at least one condition during follow-up. Restricting entry to age 40 reduces heterogeneity from early-onset disease processes, while enforcing a disease-free baseline enables modelling of incident disease acquisition from a common origin. The CTBN framework is therefore used to characterise the progression of morbidity accumulation, capturing transitions from zero conditions to one, from one to two, and so on, rather than conditioning on a fixed level of multimorbidity. Individuals with only a single condition during follow-up were retained, as each condition-specific transition contributes information on movement from preceding disease states and does not require prior multimorbidity.

Cohort derivation proceeded sequentially. Among 229,725 participants with linked UK Biobank primary-care records, 38,698 had no recorded InflAIM diagnosis and were excluded, leaving 191,027 individuals. We further excluded 137,793 participants whose first InflAIM diagnosis occurred before age 40, 15,329 with no InflAIM diagnosis at or after age 40, and 4,347 whose first recorded condition at age 40 or older could not be verified as incident, yielding a final analytic cohort of \textbf{33,558 individuals}. This cohort contributed 139,943 clinical event records spanning 1970--2017, with an observed disease-free period at the start of adult follow-up.

Follow-up begins at age 40, which is treated as the common disease-free origin. Age is incorporated as a time-varying covariate, evolving deterministically over risk intervals, while all remaining covariates (sex, ethnicity, lifestyle factors, and biomarker profiles) are treated as fixed and taken from UK Biobank baseline assessments conducted between 2006 and 2010. These baseline measurements are therefore interpreted as mid-trajectory snapshots rather than strictly pre-baseline exposures, and estimated covariate effects should be understood as associations with disease onset within this observational framework. This distinction is revisited in the results and discussion sections when interpreting covariate effects and limitations.

All inferential results are derived using the recommended CTBN spike-and-slab specification with $P = 2$ and $\theta = 1$, as identified from the simulation study, and subsequently fitted to the complete cohort. The model includes 31 fixed covariates, with age incorporated as a time-varying covariate. For visualisation, the three inflammatory conditions ([Infl],RS, [Infl],DE, [Infl],CLRD) are displayed as a contiguous block on the right-hand side of the column axis in all heat maps and are highlighted using an \textbf{amber annotation bar}, to emphasise their shared inflammatory subgroup structure within the inferred disease network. All analyses were performed using R 4.5.0.

\subsection{Sensitivity of Interaction Order and Penalty on Predictive Performance}

The mean time-dependent AUC (tdAUC) over evaluation horizons for $P \in \{1, 2, 3\}$, averaged over $\theta \in \{0.5, 1, 2\}$, shows no substantial gain from $P = 1$ to $P = 2$ and no improvement (or slight degradation) at $P = 3$, consistently across all prior families (Supplementary Figure~S10); the held-out Brier score and Poisson log-likelihood (Supplementary Table~S2) and the near-insensitivity of all priors to $\theta$ (Supplementary Figure~S1) confirm these findings. We therefore adopt the spike-and-slab prior with $P = 2$ and $\theta = 1$, motivated by the sharper discrimination between relevant and irrelevant predictors afforded by its two-component mixture. The per-condition tdAUC profile (Supplementary Figure~S11) shows that inflammatory conditions have lower individual tdAUC than cardiometabolic conditions, reflecting their more diffuse multimorbidity profiles, though their pairwise interaction terms contribute meaningfully to the $P=2$ gain over $P=1$.

\subsection{Inferred Multimorbidity Network}
\label{sec:network}

\begin{figure}[ht]
\centering
\includegraphics[width=.85\linewidth]{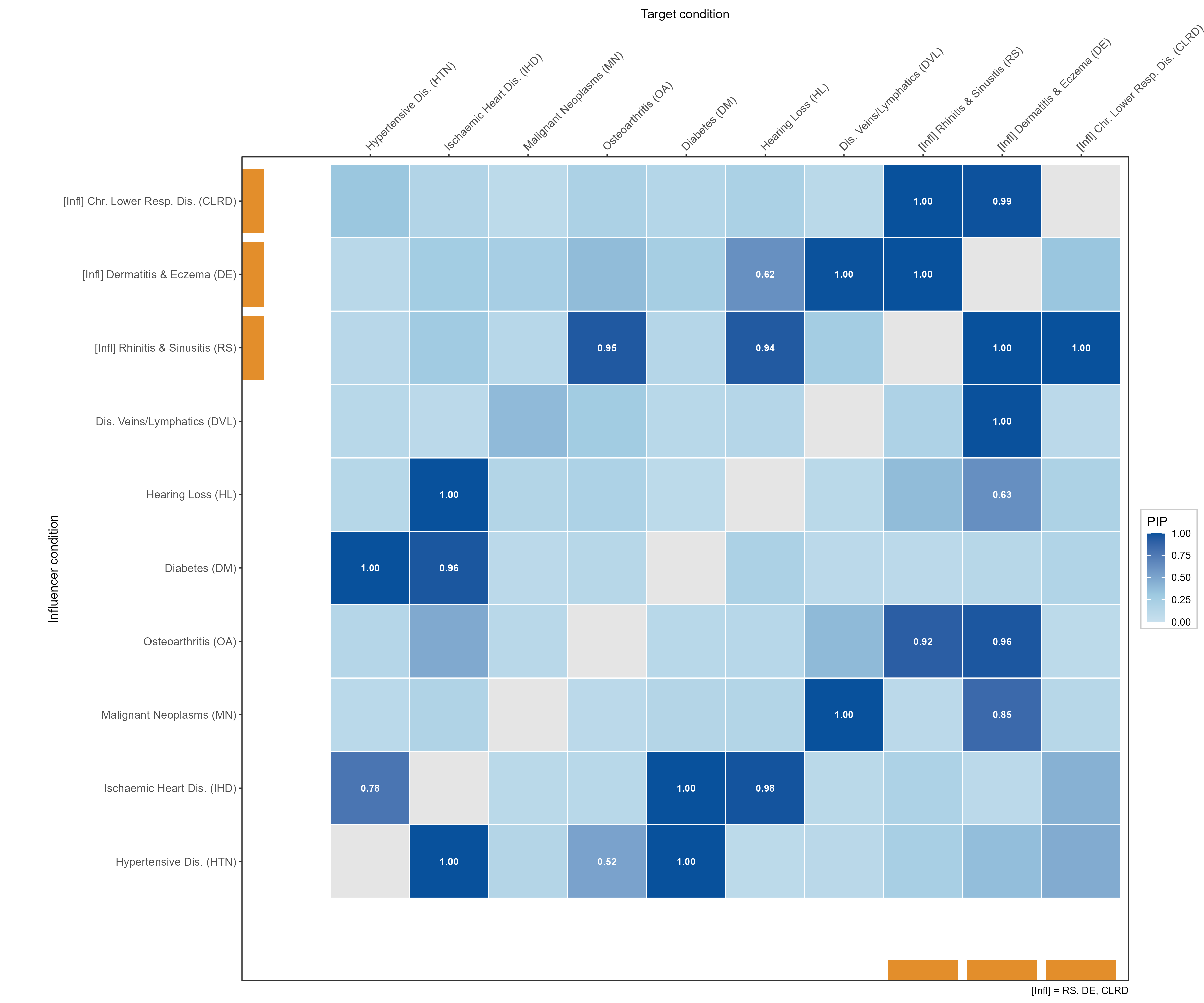}
\caption{Posterior inclusion probability matrix.  Rows = influencer; columns = target.  Bold white text: active edges (PIP\,$\geq$\,0.50).  \textbf{Amber bar} = inflammatory cluster ([Infl]\,RS, [Infl]\,DE, [Infl]\,CLRD).  Spike-and-slab;
$\theta=1$; $P=2$.} \label{fig:pip_heatmap}
\end{figure}

Figure~\ref{fig:pip_heatmap} displays the $10 \times 10$ posterior inclusion probability (PIP) heatmap of directed condition-to-condition edges. Within the cardiometabolic cluster, DM$\to$HTN, HTN$\to$IHD, IHD$\to$DM, and HL$\to$IHD all reach PIP\,$=1.00$ (with DM$\to$IHD $=0.96$ and IHD$\to$HL $=0.98$), confirming a dense, well-supported and partly bidirectional interdependence. Within the inflammatory cluster, RS$\to$DE, RS$\to$CLRD, CLRD$\to$RS (all $1.00$) and CLRD$\to$DE ($0.99$) achieve near-certain inclusion, consistent with shared atopic and airway-inflammation mechanisms. Cross-cluster edges are sparser, the most robust being MN$\to$DVL ($1.00$), OA$\to$DE ($0.96$), and OA$\to$RS ($0.92$), with HL$\to$DE ($0.63$) and HTN$\to$OA ($0.52$) weaker but non-negligible.

Supplementary Figure~S2 presents the corresponding posterior mean rate ratios for all active edges, where the posterior mean rate ratio for edge $j \to m$ is defined as $\widehat{\text{RR}}_j^m = \exp(\hat{\beta}^m_{\{j\}})$ with $\hat{\beta}^m_{\{j\}} = \mathbb{E}[\beta^m_{\{j\}} \mid \mathcal{D}]$. The largest estimated effects are $\text{DM} \to \text{HTN}$ ($\widehat{\text{RR}} = 2.13$) and $\text{HTN} \to \text{DM}$ ($\widehat{\text{RR}} = 2.61$), indicating a strongly mutually reinforcing cardiometabolic relationship. Within the inflammatory cluster, $\text{CLRD} \to \text{RS}$ ($\widehat{\text{RR}} = 2.09$) and $\text{RS} \to \text{DE}$ ($\widehat{\text{RR}} = 1.95$) represent the largest inflammatory cross-effects. 
$\text{HTN} \to \text{IHD}$ ($\widehat{\text{RR}} = 1.67$) and $\text{IHD} \to \text{DM}$ ($\widehat{\text{RR}} = 2.01$) further highlight the magnitude of cardiometabolic cascading. All active edges are excitatory ($\widehat{\text{RR}} > 1$), 
indicating that the presence of any active influencer condition uniformly accelerates the onset of its target condition.

\begin{figure}[H]
\centering
\includegraphics[width=0.9\linewidth]{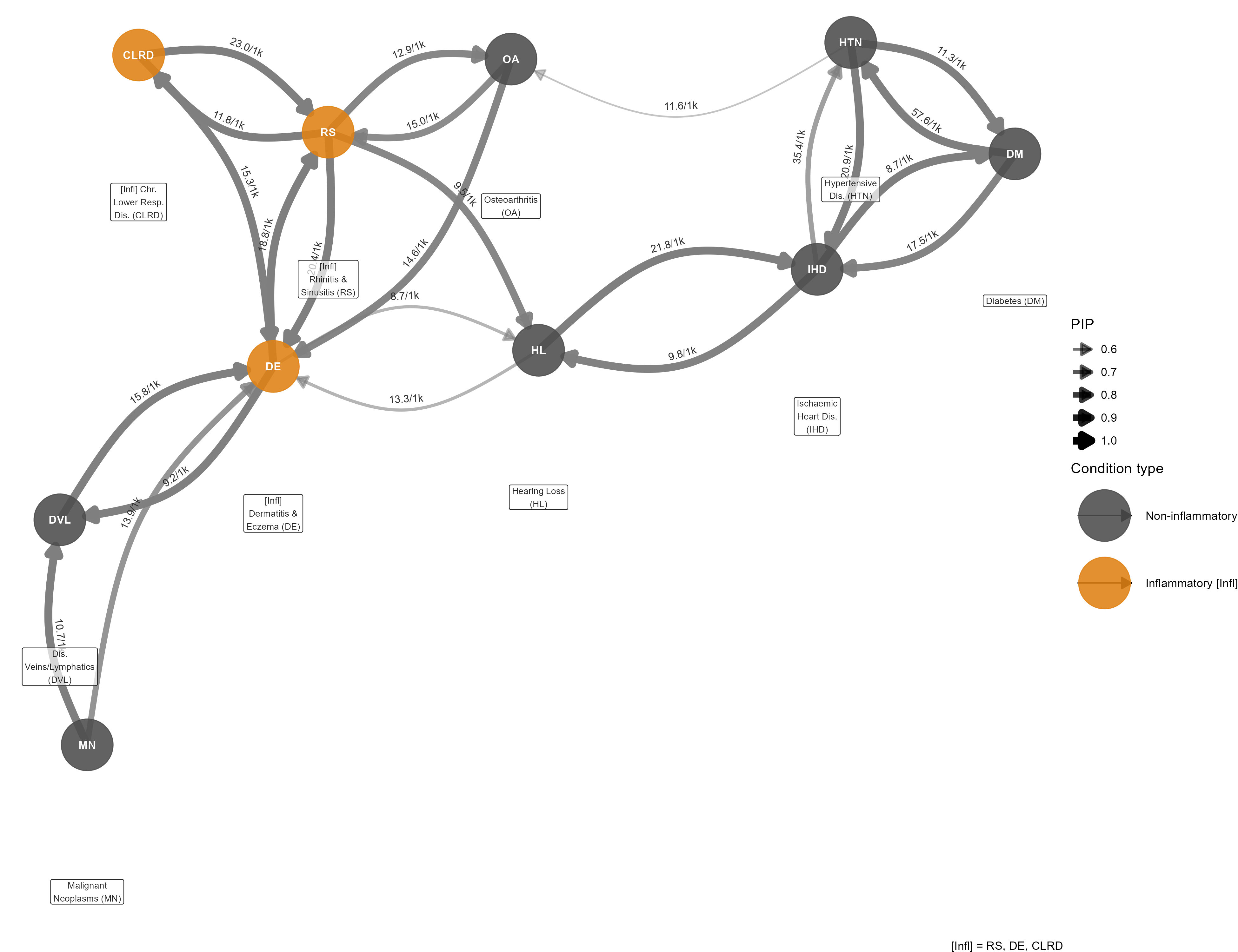}
\caption{Directed multimorbidity network (PIP\,$\geq$\,0.50).  \textbf{Orange nodes} = inflammatory conditions.  Edge width $\propto$ PIP; labels = conditional intensity (events/1,000\,py).}
\label{fig:network_graph}
\end{figure}

Figure~\ref{fig:network_graph} represents the network as a directed graph annotated with per-edge transition rates (per 1{,}000 person-years). Two distinct modules are visible: a dense cardiometabolic cluster comprising HTN, IHD, DM, and HL, and an inflammatory cluster comprising RS, DE, and CLRD, with OA, MN, and DVL occupying more peripheral roles. The 
highest transition rates are observed for $\text{DM} \to \text{HTN}$ ($57.6~\text{per}~1{,}000$) and $\text{HTN} \to \text{IHD}$ 
($35.4~\text{per}~1{,}000$), reflecting the high progression burden within the cardiometabolic pathway. Full active edge details are provided in Supplementary Table~S3.

\subsection{Conditional Risk of a Second Long-Term Condition}
\label{sec:conditional_risk}

The five-year excess cumulative incidence (Supplementary Figure~S3) maps the 5-year excess cumulative incidence $\Delta^{(1)}_m(j) = F_m(5 \mid X_j = 1) - F_m(5 \mid X_j = 0)$ for all 90 off-diagonal condition pairs, evaluated at population-mean covariate values. The largest excess risk arises within the cardiometabolic cluster: individuals with 
existing DM face a $+12.4$ percentage point (pp) excess risk of developing HTN (conditional risk $25.0\%$ vs.\ baseline $12.6\%$; $\widehat{\text{RR}} = 2.13$), representing the single most clinically consequential directed association in the network. Other notable cardiometabolic pairs include $\text{HTN} \to \text{IHD}$ ($+3.8$\,pp; $\widehat{\text{RR}} = 1.67$), $\text{HL} \to \text{IHD}$ ($+4.2$\,pp; $\widehat{\text{RR}} = 1.74$), and $\text{HTN} \to \text{DM}$ ($+3.4$\,pp; $\widehat{\text{RR}} = 2.61$), the latter notable for its large relative effect despite a lower baseline incidence of DM ($2.1\%$).

Within the inflammatory cluster, $\text{CLRD} \to \text{RS}$ ($+5.5$\,pp; $\widehat{\text{RR}} = 2.09$) and $\text{RS} \to \text{DE}$ ($+4.6$\,pp; 
$\widehat{\text{RR}} = 1.95$) confirm that chronic airway disease substantially elevates the onset of upper respiratory and atopic conditions, consistent with shared mucosal inflammatory pathways. $\text{DE} \to \text{RS}$ also contributes a non-trivial excess risk ($+3.6$\,pp; $\widehat{\text{RR}} = 1.70$), indicating bidirectional reinforcement within the atopic triad. A modest but notable cross-cluster association is observed for $\text{MN} \to \text{DVL}$ ($+2.2$\,pp; $\widehat{\text{RR}} = 1.74$), plausibly reflecting treatment-related vascular sequelae. Several entries yield near-zero or marginally negative excess risks, most prominently $\text{DVL} \to \text{HTN}$ ($-0.9$\,pp), suggesting that DVL 
confers negligible additional risk for HTN beyond shared background factors. The top 15 condition pairs by excess absolute risk are provided in Supplementary Table~S4.

\subsubsection{Synergistic Risk of a Third Long-Term Condition}
\label{sec:interaction_effects}

For a patient with two conditions $j$ and $k$ simultaneously present, the onset 
rate for a third condition $m$ decomposes as:
\[
q_m(j,k)
= \underbrace{\exp(\beta^0_m + \bar{\mathbf{Z}}^\top\boldsymbol{\gamma}_m)}_{\text{baseline}}
\times
\underbrace{e^{\beta^m_{\{j\}}}}_{\mathrm{RR}_j}
\times
\underbrace{e^{\beta^m_{\{k\}}}}_{\mathrm{RR}_k}
\times
\underbrace{e^{\beta^m_{\{j,k\}}}}_{\text{synergistic multiplier}}.
\]
The synergistic multiplier $e^{\beta^m_{\{j,k\}}}$ is the exponentiated two-way interaction coefficient of equation~\eqref{eqn1}: a value $> 1$ indicates super-multiplicative synergy and a value $< 1$ sub-multiplicative antagonism. Active interactions are declared at $\text{PIP}_{jk}^{m} \geq 0.50$.

On the cumulative incidence scale we define the \emph{synergistic excess} at horizon $\tau$ as
\begin{equation}\label{eq:synergy_excess}
\begin{aligned}
\Delta F_m(\tau)
&= \underbrace{F_m(\tau \mid X_j=1, X_k=1)}_{\text{joint exposure}}\\
&-\;
\underbrace{\bigl[F_m(\tau \mid X_j=1, X_k=0) + F_m(\tau \mid X_j=0, X_k=1) - F_m(\tau \mid X_j=0, X_k=0)\bigr]}_{\text{additive counterfactual}},
\end{aligned}
\end{equation}
evaluated at population-mean covariate values $\bar{\mathbf{Z}}$. The bracketed term is the cumulative incidence that would arise if the two conditions acted purely additively on the cumulative incidence scale (so $\Delta F_m(\tau)$ equals zero in the absence of a two-way interaction in the log-rate); $\Delta F_m(\tau)$ therefore translates the interaction coefficient $\beta^m_{\{j,k\}}$ into clinically interpretable percentage-point units. Positive $\Delta F_m(\tau)$ indicates super-multiplicative excess (joint exposure yields more cumulative risk than the sum of marginal exposures), negative $\Delta F_m(\tau)$ indicates sub-multiplicative attenuation.

Four triplets achieve high posterior inclusion ($\text{PIP}_{jk}^{m} \geq 0.997$), all exhibiting sub-multiplicative antagonism (synergistic multiplier $< 1$; Supplementary Table~S5, Figures~S4--S5). The largest absolute excesses arise in the cardiometabolic cluster: $\{\text{RS},\,\text{DM}\} \to \text{HTN}$ ($\Delta F_m(5) = 7.7$\,pp; multiplier $0.646$) and $\{\text{IHD},\,\text{DM}\} \to \text{HTN}$ ($7.4$\,pp; multiplier $0.631$), where large individual effects are attenuated on joint exposure, followed by $\{\text{HTN},\,\text{DM}\} \to \text{IHD}$ ($3.3$\,pp; multiplier $0.642$). By contrast, $\{\text{HTN},\,\text{IHD}\} \to \text{DM}$, despite the largest individual rate ratios ($\widehat{\text{RR}}_j = 2.61$, $\widehat{\text{RR}}_k = 2.01$), shows the strongest attenuation (multiplier $0.591$) and the smallest excess ($1.1$\,pp), consistent with a ceiling effect once individual risks are already elevated. The joint cumulative-incidence trajectories lie below the multiplicative combination of the marginal trajectories in every case, yet remain substantially above baseline, most strikingly for $\{\text{IHD},\,\text{DM}\} \to \text{HTN}$, which approaches $50\%$ by year~10.

\subsection{Covariate Associations}
\label{sec:covariates_results}

Covariate posterior inclusion probabilities (Supplementary Figure~S12) show that age achieves PIP\,=\,1.00 across all ten conditions, confirming it as the strongest and most consistent predictor of onset. Among fixed markers, WBC, platelet, GlycA, neutrophil, and lymphocyte counts attain the highest PIPs for cardiometabolic targets, consistent with EHR evidence linking systemic inflammatory markers to cardiovascular risk; current and former smoking are selected for [Infl]\,CLRD; BMI\,$>$\,30 for HTN, MN, OA, DM, and HL; and sex for IHD, OA, DM, HL, [Infl]\,RS and [Infl]\,CLRD.

\begin{figure}[H]
\centering
\includegraphics[width=\linewidth]{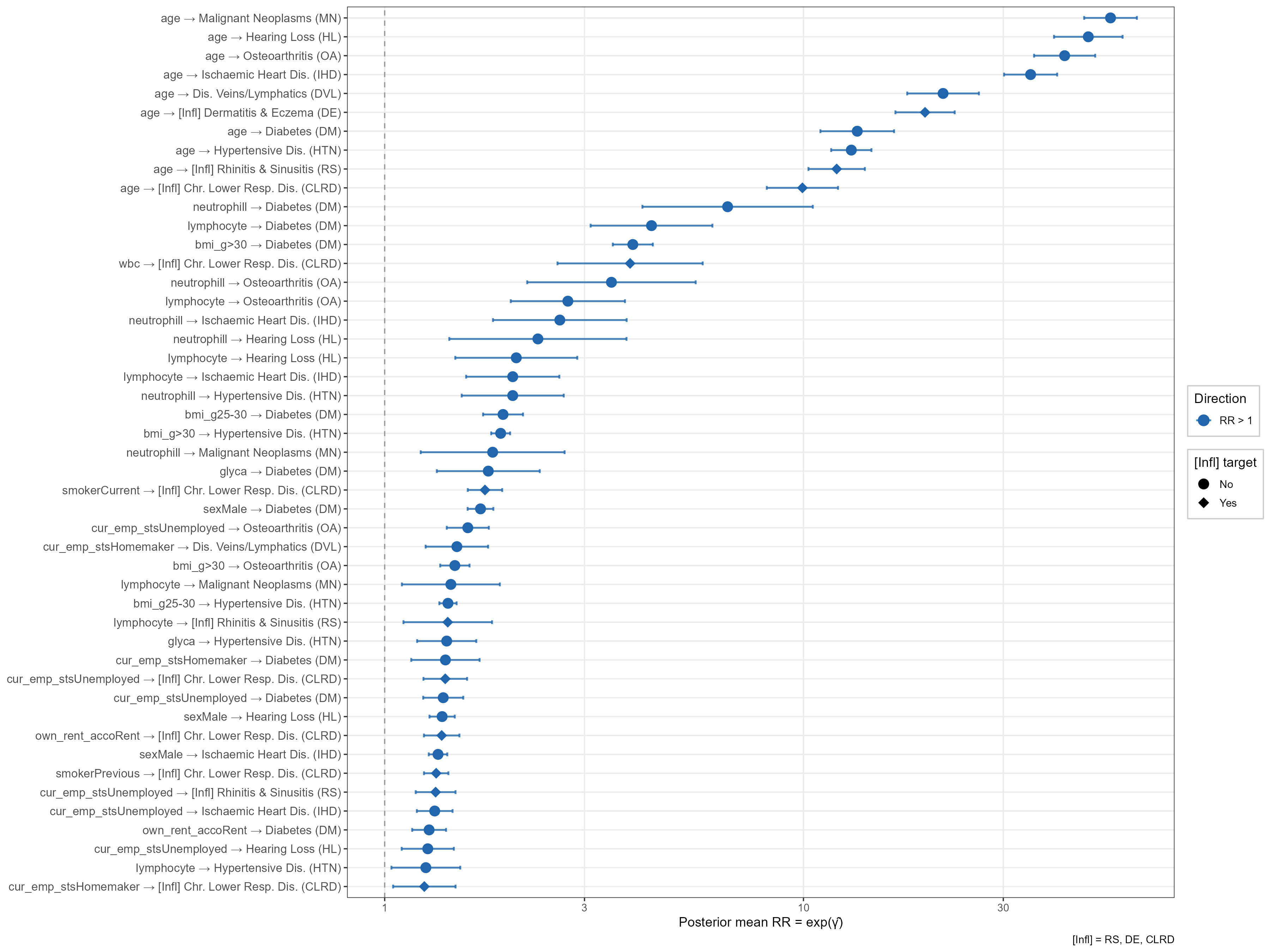}
\caption{Forest plot of posterior mean rate ratios for excitatory covariates (RR $>1$ and PIP\,$\geq$\,0.50).  95\% credible intervals from NUTS.
Spike-and-slab; $\theta=1$; $P=2$.}
\label{fig:rr_covariates_forest}
\end{figure}

Posterior mean rate ratios (Figure~\ref{fig:rr_covariates_forest}) confirm clinically expected directions across excitatory covariates: older age elevates onset across all ten conditions; male sex raises rates for IHD, DM, HL, and CLRD; current and previous smoking substantially elevate CLRD; elevated GlycA accelerates HTN and DM; higher BMI ($>$30) raises HTN, DM, and OA; elevated neutrophil and lymphocyte counts are broadly associated with higher rates across cardiometabolic, inflammatory, and neoplastic conditions, with WBC additionally contributing to CLRD; and socioeconomic indicators (unemployment, renting) elevate several conditions including IHD, OA, DM, HL and RS. Conversely, a distinct set of inhibitory covariates is associated with reduced onset (Supplementary Figure~S7): albumin is the most consistent protective factor across all ten conditions, platelet count and WBC reduce rates for DM, IHD, MN, and DVL, and Asian and European ethnicities show condition-specific reductions relative to the African reference. The full covariate rate-ratio heatmap is given in Supplementary Figure~S6.

\section{Discussion}
\label{sec:discussion}

\subsection{Summary of Main Findings}
Simulation studies confirm that the spike-and-slab achieves the best 
RMSE, FPR control, coverage calibration, and Poisson log-likelihood among the 
four priors evaluated, at the cost of reduced sensitivity (TPR rank 4 across 
all analysis models); the three continuous shrinkage priors (structured normal, Bayesian LASSO, and regularised horseshoe), by contrast, cannot perform hard variable selection at a fixed $0.5$ threshold, declaring almost all coefficients active (TPR $\approx$ FPR $\approx 100\%$) and retaining only modest, above-random ranking ability (selection AUC $\approx 68\text{--}71\%$), versus $\approx 85\text{--}95\%$ for the spike-and-slab.

Applied to UK Biobank primary-care EHRs, the recommended spike-and-slab model ($P = 2$, $\theta = 1$) recovers sparse, clinically plausible networks of multimorbidity progression organised into two dominant modules: a cardiometabolic cluster with strongly bidirectional edges among DM, HTN, and IHD (DM$\to$HTN: $\widehat{\text{RR}} = 2.13$, $+12.4$\,pp excess 5-year risk; HTN$\to$DM: $\widehat{\text{RR}} = 2.61$), and an inflammatory cluster of dense within-cluster edges among RS, DE, and CLRD (CLRD$\to$RS: $\widehat{\text{RR}} = 2.09$; RS$\to$DE: $\widehat{\text{RR}} = 1.95$), joined by sparse but robust cross-cluster links (OA$\to$DE, PIP\,$=0.96$; OA$\to$RS, PIP\,$=0.92$). All four high-inclusion triplets ($\text{PIP}_{jk}^{m} \geq 0.997$) exhibit sub-multiplicative antagonism concentrated in the cardiometabolic cluster, most prominently $\{\text{RS}, \text{DM}\} \to \text{HTN}$ ($\Delta F_m(5) = 7.7$\,pp; multiplier $0.646$) and $\{\text{IHD}, \text{DM}\} \to \text{HTN}$ ($7.4$\,pp; multiplier $0.631$).

\subsection{Comparison with Related Methodological Work}

Relative to \citet{faruqui2021functional}, who used frequentist $L_1$-regularisation with cross-validated penalty selection, our framework provides full posterior uncertainty quantification (credible intervals and PIPs) and enables joint estimation of main effects and pairwise interactions under order-dependent shrinkage; simulation confirms that the spike-and-slab attains TPR $\approx 73\%$ at FPR $\approx 4\%$ for interactions under correct specification, a marked improvement in precision--recall over continuous shrinkage alternatives that collapse to near-$100\%$ TPR and FPR.

Compared with the non-parametric Bayesian CTBN of \citet{yang2016learning, lasserre2021constraint}, our parametric log-linear approach favours interpretability and computational tractability, which is preferable in the large-cohort EHR setting; a non-parametric formulation would be warranted only if exploratory analysis indicated substantial non-linearity in the intensity--covariate relationship.

The comparison of our four priors provides nuanced insights aligned with 
theoretical predictions. The dominance of spike-and-slab for parameter recovery, 
FPR control, and coverage calibration corroborates \citet{bhadra2019lasso}, who 
showed that explicit zero-probability spike components outperform continuous 
shrinkage for variable selection under moderate-to-high true sparsity. The behaviour of the horseshoe prior, characterised by a highly random but moderate selection accuracy (AUC $\approx 68\text{--}70\%$) in combination with the inclusion of almost all parameters at the $0.5$ threshold (TPR $\approx$ FPR $\approx 100\%$) aligns with the observation of \citet{piironen2017sparsity} that the regularised horseshoe controls the complexity of the effective model at a global level but does not, in itself, produce posterior variable selection at the level of individual coefficients without an additional thresholding mechanism. In our simulation study, the resulting pseudo-PIP ranking encodes real but limited information about variable importance; however, virtually no coefficients are shrunk below the inclusion cutoff. The structured normal prior and the Bayesian LASSO exhibit qualitatively similar behaviour. The competitive predictive performance of all four priors under correct specification is consistent with the well-known insensitivity of predictive accuracy to prior specification \citep{VehtariGelmanGabry2017}, and is confirmed here by the near-identical Poisson log-likelihood and Brier scores across priors within each analysis model (Table~\ref{tab:pred_metrics}).

\subsection{Comparison with Related Work on Disease Networks}

The recovered cardiometabolic structure is broadly consistent with large-scale multimorbidity studies. \citet{barnett2012epidemiology} identified diabetes, hypertension, and ischaemic heart disease as a persistent cluster in Scottish primary care; our dynamic model adds directionality and quantifies the progression burden, with diabetes the most consequential existing condition, accelerating both hypertension ($\widehat{\text{RR}} = 2.13$, $+12.4$\,pp at 5 years) and IHD onset ($\widehat{\text{RR}} = 1.40$). The implied temporal ordering aligns with the chronological atlas of \citet{kuan2019chronological}, in which diabetes typically precedes hypertension and IHD, while the bidirectionality of the DM--HTN relationship (HTN$\to$DM: $\widehat{\text{RR}} = 2.61$) reflects shared mechanisms, including renin--angiotensin activation and glucocorticoid-mediated insulin resistance, that propagate risk in both directions.

The inflammatory cluster extends this picture in ways cross-sectional analyses have not quantified. The bidirectional edges CLRD$\to$RS ($\widehat{\text{RR}} = 2.09$, $+5.5$\,pp) and RS$\to$CLRD ($\widehat{\text{RR}} = 1.79$, $+2.5$\,pp) indicate dense reinforcement between upper and lower airway disease, consistent with the atopic-march hypothesis in which chronic upper-airway inflammation predisposes to lower-airway disease through shared eosinophilic and Th2-mediated mechanisms \citep{cheng2024chronic, valabhji2024prevalence}; RS$\to$DE ($\widehat{\text{RR}} = 1.95$, $+4.6$\,pp) reflects the atopic triad of rhinitis, eczema and asthma \citep{yang2024longitudinal}. The cross-cluster edge OA$\to$RS ($\widehat{\text{RR}} \approx 1.36$, PIP\,$= 0.92$) is less well characterised; one plausible mechanism is shared neuro-inflammatory sensitisation, though this warrants further investigation.

The sub-multiplicative interactions carry direct clinical implications. All four active triplets involve conditions whose individual onset rates are already elevated, and their multipliers (range $0.591$--$0.646$) indicate joint effects that, while well above baseline, are attenuated relative to the naive multiplicative expectation, most strikingly for $\{\text{HTN}, \text{IHD}\} \to \text{DM}$ (smallest excess, $1.1$\,pp). This is consistent with evidence that the components of metabolic syndrome act through overlapping pathways, so that each additional condition contributes less once the shared mechanisms are saturated \citep{valabhji2024prevalence}.

\subsection{Limitations}

Several limitations merit consideration. First, the log-linear CTBN framework assumes piecewise-constant transition intensities over at-risk intervals, thereby precluding smoothly varying hazard functions; non-parametric extensions could relax this constraint, albeit at increased computational cost. Second, the absorbing-state assumption may be violated for conditions characterised by relapsing–remitting trajectories (e.g. rhinitis, dermatitis). Third, the set of ten conditions used for demonstration was selected primarily on the basis of prevalence and is therefore not optimised from a clinical standpoint: some conditions (notably hearing loss) are strongly age-dependent and may enter the network predominantly through shared determinants such as biological ageing and socioeconomic deprivation rather than through direct pathophysiological relationships, while several prototypical inflammatory diseases (for example, rheumatoid arthritis) fall below the prevalence threshold and are consequently excluded. As a result, well-established associations, such as that between rheumatoid arthritis and ischaemic heart disease, cannot be recovered in the current analysis; a more clinically curated condition set is thus a priority for the full InflAIM study. 

Fourth, non-age covariates are fixed in the UK Biobank baseline assessment (2006–2010), rather than standardised to age 40, so that for some participants specific biomarkers were measured after the onset of one or more conditions. Covariate effects should therefore be interpreted as marker–outcome associations within the cohort, rather than as strictly pre-onset risk-factor effects. Fifth, the UK Biobank cohort is not fully representative of the general UK population due to healthy volunteer bias; therefore, extrapolation of absolute incidence estimates to population-level projections should be undertaken with caution. Sixth, the findings of the sub-multiplicative interaction, although strongly statistically supported (PIP $\ge$ 0.997), require replication in independent cohorts before informing clinical decision-making, as the estimates of the corresponding interaction parameters are sensitive to misspecification of the underlying multiplicative baseline model. 

Seventh, the time-dependent AUC uses the inverse-probability-of-censoring weighted estimator of \citet{BlancheCommengesJacqmin-Gadda2013}; because the CTBN is itself generative for event times, a fully Bayesian alternative that imputes censored event times from the posterior predictive distribution, obviating IPCW weights and propagating posterior uncertainty into the AUC, is a natural direction for future methodological development.


\subsection{Conclusion and recommendations}
\label{sec:conclusion}

This study developed and assessed a structured Bayesian continuous-time Bayesian network (CTBN) framework, incorporating four order-dependent shrinkage priors, to model multimorbidity trajectories in large-scale electronic health records. Across all three analysis models, the spike-and-slab prior with $P = 2$ and $\theta = 1$ is recommended as the main specification for network learning: it delivers the best RMSE, FPR control, coverage calibration, and Poisson log-likelihood, together with the sharpest discrimination between active and inactive parameters (selection AUC $\approx 85\text{--}95\%$ vs.\ $\approx 68\text{--}71\%$ for the continuous priors), and is robust to order misspecification. The regularised horseshoe and Bayesian LASSO are not recommended for hard selection, as they classify nearly all coefficients as active; the structured normal is a viable secondary option where hard selection is not required, and the continuous priors may remain useful for prediction when point estimation rather than structure recovery is the goal.

Applied to the UK Biobank cohort, the framework yields a statistically robust and clinically plausible network that reproduces established cardiometabolic and atopic pathways while quantifying their effect sizes, directionality, and higher-order interaction structure. Extending it to the full set of 60 InflAIM conditions constitutes a substantial advance in population-scale disease-network modelling, with translational implications for prevention, risk stratification, and precision medicine in complex multimorbidity.

\section*{Conflicts of interest}
The authors declare that they have no competing interests.

\section*{Funding}
This work was supported by NIHR grant NIHR205461 as part of the InflAIM programme. The views expressed are those of the authors and not necessarily those of the NIHR or the Department of Health and Social Care.

\section*{Code and Data availability}
All Stan model files for the four shrinkage priors, R scripts for the EHR-to-interval conversion, the simulation pipeline (analysis models A1--A3), and plotting code for every figure and table in the paper and Supplementary Material are available at \url{https://github.com/Inflaim-Study/CTBN-Multimorbidity-Paper}. This research used the UK Biobank Resource under Application Number 303849. UK Biobank primary-care data are not publicly available but can be accessed by approved researchers through the UK Biobank Access Management System (\url{https://www.ukbiobank.ac.uk/}).

\section*{Author contributions statement}

O.R.O. conceived the study, developed the methodology, conducted the formal analysis and investigation, curated the data, developed the software, validated the results, created the visualizations, and wrote the original draft and reviewed the manuscript. S.S.P. and M.K. extracted and prepared the UK Biobank analysis dataset. A.J.M. and A.L. supervised the project, provided resources, administered the project, acquired funding, and reviewed the manuscript.

\section*{Acknowledgments}
This work was supported by NIHR grant NIHR205461 (InflAIM programme) and by the MRC Better Methods, Better Research grant UKRI491; the views expressed are those of the authors and not necessarily those of the NIHR or the Department of Health and Social Care. We thank the anonymous reviewers, the PPIE group members and clinical experts, and the whole InflAIM team for their contributions.

\bibliographystyle{apalike}
\bibliography{reference}

\clearpage


\section*{Supplementary material}

# The InflAIM Condition Set

Table S1 lists the full set of 60 inflammatory, autoimmune, and metabolic long-term conditions used to define the *healthy-at-40* cohort entry criterion, grouped by disease category. The ten conditions modelled as outcomes in the primary CTBN trajectory analysis (the ten most prevalent within the final analytic cohort, $n = 33{,}558$) are shaded and marked with ✓; the remaining 50 conditions contribute only to the cohort-entry definition and are not modelled as outcomes.

Table S1: Full list of the 60 InflAIM long-term conditions included in the UK Biobank cohort definition, grouped by disease category. Conditions included in the primary CTBN trajectory analysis (the ten most prevalent) are shaded and marked with ✓.

| No. | Condition | Primary[a] |
|---|---|---|
| **1. Cardiovascular and Circulatory** | | |
| 1 | Hypertensive diseases | ✓ |
| 2 | Ischaemic heart diseases | ✓ |
| 3 | Other forms of heart disease | |
| 4 | Chronic rheumatic heart diseases | |
| 5 | Diseases of arteries, arterioles and capillaries | |
| 6 | Diseases of veins, lymphatic vessels and lymph nodes, not elsewhere classified | ✓ |
| 7 | Cerebrovascular disease | |
| **2. Metabolic and Endocrine** | | |
| 8 | Diabetes | ✓ |
| 9 | Disorders of thyroid gland | |
| 10 | Disorders of bone density and structure | |
| 11 | Hyperparathyroidism | |
| 12 | Hypoparathyroidism | |
| 13 | Addison's disease | |
| **3. Respiratory** | | |
| 14 | Chronic lower respiratory diseases | ✓ |
| 15 | Rhinitis and sinusitis | ✓ |
| 16 | Fibrosing alveolitis / unspecified alveolitis | |
| **4. Neoplasms** | | |
| 17 | Malignant neoplasms | ✓ |
| 18 | Malignant neoplasms of lymphoid, haematopoietic and related tissue | |
| 19 | In situ neoplasms | |
| **5. Musculoskeletal and Connective Tissue** | | |
| 20 | Osteoarthritis | ✓ |
| 21 | Inflammatory polyarthropathies | |
| 22 | Ankylosing spondylitis | |
| 23 | Systemic connective tissue disorders | |
| 24 | Spinal degenerative disease | |
| **6. Gastrointestinal** | | |

*Continued on next page*

*Supplementary Table S1 (continued)*

| No. | Condition | Primary[a] |
|---|---|---|
| 25 | Diseases of oesophagus, stomach and duodenum | |
| 26 | Irritable bowel syndrome | |
| 27 | Diverticular disease of the intestine | |
| 28 | Noninfective enteritis and colitis | |
| 29 | Disorders of gallbladder, biliary tract and pancreas | |
| 30 | Diseases of the liver | |
| **7. Neurological and Mental Health** | | |
| 31 | Epilepsy | |
| 32 | Migraine | |
| 33 | Mood disorders | |
| 34 | Neurotic, stress-related and somatoform disorders | |
| 35 | Polyneuropathies and other disorders of the peripheral nervous system | |
| 36 | Dementia | |
| 37 | Parkinson disease | |
| 38 | Multiple sclerosis (and optic neuritis) | |
| 39 | Eating disorders | |
| 40 | Sleep disorders | |
| 41 | Schizophrenia | |
| 42 | Drug or alcohol misuse | |
| **8. Dermatological** | | |
| 43 | Dermatitis and eczema | ✓ |
| 44 | Psoriasis | |
| **9. Renal and Urological** | | |
| 45 | Chronic kidney disease | |
| 46 | Disorders of the prostate | |
| **10. Haematological** | | |
| 47 | Iron deficiency anaemia | |
| **11. Ophthalmological and ENT** | | |
| 48 | Glaucoma | |
| 49 | Cataract | |
| 50 | Other retinal disorders (macular degeneration) | |
| 51 | Visual disturbances and blindness | |
| 52 | Retinal detachments and breaks | |
| 53 | Conductive and sensorineural hearing loss | ✓ |
| 54 | Meniere disease | |
| **12. Gynaecological** | | |
| 55 | Inflammatory diseases of the female pelvic organs | |
| 56 | Endometriosis | |
| **13. Infectious** | | |
| 57 | Tuberculosis | |
| 58 | Chronic viral hepatitis | |
| 59 | Other spirochaetal infections | |
| 60 | HIV disease | |

[a] **Primary analysis:** ✓ indicates that the condition is among the ten most prevalent within the final analysis cohort ($n = 33{,}558$) and is modelled as an outcome in the primary CTBN trajectory model; shaded rows identify these ten conditions. The remaining 50 conditions are used solely to define the healthy-at-40 cohort-entry criterion and are not modelled as outcomes. *Data source:* UK Biobank primary care (GP) clinical event records, with conditions identified using Read codes mapped to the InflAIM framework. Cohort entry requires participants to be free of all 60 conditions at age 40; follow-up begins at age 40 and is censored at the last observed clinical event.

# S1 Detailed Specification of the Four Shrinkage Priors

This section gives the full hierarchical specification, marginal density, and closed-form full conditionals for each of the four shrinkage priors compared in the main text. Throughout, $\beta_j$ denotes a coefficient with associated interaction order $p_j \in \{1, \dots, P\}$ (with $p_j = 1$ for main effects and $p_j = r$ for an $r$-way interaction), $\theta > 0$ is the order-penalty hyperparameter, and $\boldsymbol{\gamma}_m$ denotes the covariate coefficients which are not subject to order-based penalisation. We write $n_p = |\{j : p_j = p\}|$ for the number of coefficients of order $p$, and $n_m$ for the effective sample size for condition $m$ (number of at-risk intervals).

## S1.1 Hierarchical Structured (HS) Prior

The hierarchical structured prior places an order-specific inverse-gamma hyperprior on the slab variance.

### S1.1.1 Hierarchy

$$\beta_j \mid \sigma^2_{p_j} \sim \mathcal{N}\Big(0,\ \sigma^2_{p_j}\, e^{-\theta p_j}\Big), \qquad j = 1, \dots, J, \tag{S1}$$

$$\sigma^2_p \sim \frac{1}{\Gamma(p+1)}\,\text{Inv-Gamma}(a_0, b_0), \qquad p = 0, 1, \dots, P, \tag{S2}$$

$$\gamma_k \mid \sigma^2_\gamma \sim \mathcal{N}(0, \sigma^2_\gamma), \qquad k = 1, \dots, K, \tag{S3}$$

$$\sigma^2_\gamma \sim \text{Inv-Gamma}(a_0, b_0). \tag{S4}$$

The factorial scaling term $\Gamma(p+1)^{-1}$ in the hyperprior accounts for the combinatorially increasing number of possible $p$-way interaction terms and induces an additional level of penalisation for higher-order interactions beyond the exponential decay $e^{-\theta p_j}$. In combination, these two mechanisms ensure that main effects are subjected to only mild regularisation, whereas higher-order interaction terms are strongly shrunk towards zero. The hyperparameters are commonly chosen as $a_0 = b_0 = 0.001$ in order to concentrate the mass of the hyperprior in a neighbourhood of zero.

### S1.1.2 Marginal density

Marginalising over $\sigma^2_{p_j}$ yields a scaled Student-$t$ marginal:

$$p(\beta_j) = \frac{\Gamma(a_0 + \frac{1}{2})}{\Gamma(a_0)\sqrt{2\pi b_0\,\Gamma(p_j+1)\,e^{\theta p_j}}}\left(1 + \frac{\beta_j^2}{2b_0\,\Gamma(p_j+1)\,e^{\theta p_j}}\right)^{-(a_0+\frac{1}{2})}, \tag{S5}$$

a $t_{2a_0}$ distribution with scale that shrinks with interaction order. The heavy-tailed marginal allows individual coefficients to escape shrinkage when strongly supported by data.

### S1.1.3 Full conditionals

$$\sigma_p^2 \mid \boldsymbol{\beta} \sim \text{Inv-Gamma}\left(a_0 + \frac{n_p}{2},\;\; b_0 + \frac{e^{\theta p}}{2} \sum_{j:\, p_j = p} \beta_j^2\right), \tag{S6}$$

$$\sigma_\gamma^2 \mid \boldsymbol{\gamma} \sim \text{Inv-Gamma}\left(a_0 + \frac{K}{2},\;\; b_0 + \frac{1}{2} \sum_{k=1}^{K} \gamma_k^2\right). \tag{S7}$$

The full conditional distribution of $\beta_j$ is non-conjugate due to the Poisson likelihood and is therefore sampled using the No-U-Turn Sampler (NUTS). The slab variance $\sigma_{p_j}^2$ is specific to the order $p_j$, so the prior does not induce a distinct scale parameter at the individual coefficient level. Consequently, the posterior shrinkage factor employed in the pseudo-PIP construction is given by

$$\kappa_j = \left(1 + \mathcal{I}_j\, \sigma_{p_j}^2\, e^{-\theta p_j}\right)^{-1},$$

where $\mathcal{I}_j$ denotes the per-coefficient Poisson Fisher information as defined in Section S1.6. Thus, any within-order variation in $\kappa_j$ across coefficients arises solely through $\mathcal{I}_j$ (with coefficients having greater exposure-weighted data support experiencing less shrinkage). This property is what renders the structured prior suitable for selection at the level of individual coefficients.

## S1.2 Bayesian LASSO

The Bayesian LASSO (Park and Casella, 2008) induces $L_1$-type regularisation through a scale mixture of normals representation.

### S1.2.1 Hierarchy

$$\beta_j \mid \tau_j^2 \sim \mathcal{N}\left(0,\; \tau_j^2\, e^{-\theta p_j}\right), \qquad j = 1, \ldots, J, \tag{S8}$$

$$\tau_j^2 \mid \lambda_{p_j}^2 \sim \text{Exp}\left(\frac{\lambda_{p_j}^2}{2}\right), \tag{S9}$$

$$\lambda_p^2 \sim \text{Inv-Gamma}(a_0, b_0), \qquad p = 0, 1, \ldots, P, \tag{S10}$$

$$\gamma_k \mid \lambda_\gamma^2 \sim \text{Laplace}(0, \sqrt{\lambda_\gamma^2}), \qquad k = 1, \ldots, K, \tag{S11}$$

$$\lambda_\gamma^2 \sim \text{Inv-Gamma}(a_0, b_0). \tag{S12}$$

The per-order penalty $\lambda_p^2$ controls the overall degree of shrinkage for coefficients of order $p$.

### S1.2.2 Marginal density

Marginalising over $\tau_j^2$ yields the double-exponential (Laplace) marginal:

$$p(\beta_j \mid \lambda_{p_j}) = \frac{\lambda_{p_j}\, e^{\theta p_j/2}}{2} \exp\left(-\lambda_{p_j}\, e^{\theta p_j/2}\, |\beta_j|\right), \tag{S13}$$

so the effective Laplace scale shrinks exponentially with order.

### S1.2.3 Full conditionals

$$\tau_j^{-2} \mid \beta_j, \lambda_{p_j} \sim \text{Inv-Gaussian}\left(\frac{\lambda_{p_j}\, e^{\theta p_j/2}}{|\beta_j|},\;\; \lambda_{p_j}^2\, e^{\theta p_j}\right), \tag{S14}$$

$$\lambda_p^2 \mid \{\tau_j^2 : p_j = p\} \sim \text{Inv-Gamma}\left(a_0 + n_p,\;\; b_0 + \frac{1}{2}\sum_{j:\, p_j = p} \tau_j^2\right), \tag{S15}$$

$$\lambda_\gamma^2 \mid \boldsymbol{\gamma} \sim \text{Inv-Gamma}\left(a_0 + K,\;\; b_0 + \frac{1}{2}\sum_{k=1}^{K} \gamma_k^2\right). \tag{S16}$$

The Laplace marginal $\beta_j \sim \text{DE}(0, s_{p_j})$ with scale $s_{p_j} = \sqrt{\lambda_{p_j}^2}\, e^{-\theta p_j/2}$ has prior variance $v_j = 2s_{p_j}^2 = 2\lambda_{p_j}^2\, e^{-\theta p_j}$. The shrinkage factor used for the pseudo-PIP construction (Section 3.3 of the main text; derivation in Section S1.6) is therefore $\kappa_j = (1 + 2\,\mathcal{I}_j\, \lambda_{p_j}^2\, e^{-\theta p_j})^{-1}$, with $\mathcal{I}_j$ the per-coefficient Poisson Fisher information. As for the structured prior, the per-order penalty $\lambda_{p_j}^2$ carries no coefficient-level scale, so per-coefficient variation in $\kappa_j$ enters through $\mathcal{I}_j$.

## S1.3 Regularised Horseshoe

The regularised horseshoe (Piironen and Vehtari, 2017) employs a global–local shrinkage structure with a slab component that caps the local scale.

### S1.3.1 Hierarchy

$$\beta_j = z_j\, \tilde{\lambda}_j\, \tau_j, \qquad z_j \sim \mathcal{N}(0, 1), \tag{S17}$$

$$\tilde{\lambda}_j^2 = \frac{c^2 \lambda_j^2}{c^2 + \tau_j^2 \lambda_j^2}, \tag{S18}$$

$$\tau_j = \tau_0\, e^{-\theta p_j}, \tag{S19}$$

$$\lambda_j \sim \text{Cauchy}^+(0, 1), \tag{S20}$$

$$\tau_0 \sim \text{Cauchy}^+(0, \tau_{\text{scale}}), \tag{S21}$$

$$c^2 \sim \text{Inv-Gamma}\left(\frac{\nu_{\text{slab}}}{2},\; \frac{\nu_{\text{slab}}\, s_{\text{slab}}^2}{2}\right), \tag{S22}$$

$$\gamma_k \sim \mathcal{N}(0, 1). \tag{S23}$$

The non-centred parameterisation $\beta_j = z_j \tilde{\lambda}_j \tau_j$ improves HMC geometry when $\tau_j$ is small. The slab variance $c^2$ caps the effective local scale at $\sqrt{c^2}$. Defaults are $\nu_{\text{slab}} = 4$, $s_{\text{slab}} = 2$, and $\tau_{\text{scale}} = p_0/J \cdot n_m^{-1/2}$ where $p_0$ is a prior guess at the number of non-zero coefficients.

### S1.3.2 Auxiliary representation and full conditionals

The half-Cauchy distributions on $\lambda_j$ and $\tau_0$ admit an inverse-gamma scale-mixture representation. Introducing auxiliary variables $\nu_j > 0$ and $\xi > 0$:

$$\lambda_j^2 \mid \nu_j \sim \text{Inv-Gamma}\left(\tfrac{1}{2},\ \tfrac{1}{\nu_j}\right), \quad \nu_j \sim \text{Inv-Gamma}\left(\tfrac{1}{2},\ 1\right), \tag{S24}$$

$$\tau_0^2 \mid \xi \sim \text{Inv-Gamma}\left(\tfrac{1}{2},\ \tfrac{1}{\xi}\right), \quad \xi \sim \text{Inv-Gamma}\left(\tfrac{1}{2},\ 1\right). \tag{S25}$$

The closed-form full conditionals for the auxiliary variables are

$$\lambda_j^{-2} \mid \beta_j, \tau_j, \nu_j \sim \text{Inv-Gamma}\left(1,\ \frac{1}{\nu_j} + \frac{\beta_j^2}{2\tau_j^2}\right), \tag{S26}$$

$$\nu_j^{-1} \mid \lambda_j \sim \text{Inv-Gamma}\left(1,\ 1 + \frac{1}{\lambda_j^2}\right), \tag{S27}$$

$$\tau_0^{-2} \mid \{\beta_j\}, \{\lambda_j\}, \xi \sim \text{Inv-Gamma}\left(\frac{J+1}{2},\ \frac{1}{\xi} + \frac{1}{2}\sum_{j=1}^{J} \frac{\beta_j^2}{\lambda_j^2\, e^{-2\theta p_j}}\right), \tag{S28}$$

$$\xi^{-1} \mid \tau_0 \sim \text{Inv-Gamma}\left(1,\ 1 + \frac{1}{\tau_0^2}\right). \tag{S29}$$

The posterior shrinkage factor for the pseudo-PIP construction is $\kappa_j = (1+\mathcal{I}_j\, \tilde{\lambda}_j^2\, \tau_j^2)^{-1}$, where $\mathcal{I}_j$ is the per-coefficient Poisson Fisher information (Section S1.6). This is the original local–global form of Piironen and Vehtari (2017), $\kappa_j = (1 + (n/\sigma^2)\, \tau^2 \tilde{\lambda}_j^2)^{-1}$, with the homogeneous Gaussian data-precision $n/\sigma^2$ replaced by the exposure-weighted Poisson information $\mathcal{I}_j$; $\kappa_j$ is per-coefficient through both $\tilde{\lambda}_j$ and $\mathcal{I}_j$.

## S1.4 Continuous Spike-and-Slab

The spike-and-slab prior implements explicit probabilistic variable selection by combining a diffuse "slab" component with a highly concentrated "spike" component. To facilitate efficient posterior inference via Hamiltonian Monte Carlo (HMC), we introduce a continuous relaxation of this prior specification.

### S1.4.1 Hierarchy

For the interaction and main-effect coefficients, each $\beta_j$ follows a two-component normal mixture distribution with a *deterministic*, order-dependent prior inclusion probability. Writing the mixture in marginalised form,

$$\beta_j \sim \pi_{p_j}\, \mathcal{N}\left(0,\ \sigma_{p_j}^2\right) + (1 - \pi_{p_j})\, \mathcal{N}(0,\ \epsilon), \qquad j = 1, \ldots, J, \tag{S30}$$

$$\pi_p = \pi_0\, e^{-\theta p}, \qquad \pi_0 \in (0,1), \tag{S31}$$

$$\sigma_p^2 \sim \text{Inv-Gamma}(a_0, b_0), \qquad p = 0, 1, \ldots, P. \tag{S32}$$

The spike variance $\epsilon$ is specified as a small fixed constant satisfying $\epsilon \ll \sigma_{p_j}^2$, such that the continuous relaxation provides a close approximation to the discrete spike-and-slab prior.

The ordering penalty is introduced via the inclusion probability $\pi_{p_j}$ rather than through the slab variance, while the slab variance $\sigma^2_{p_j}$ is order-specific through its inverse-gamma hyperprior. In the likelihood, the mixture is evaluated on the marginalised density using Stan's numerically stable `log_mix` function, which is equivalent to

$$\text{log_sum_exp}\Big(\log \pi_{p_j} + \log \mathcal{N}(\beta_j \mid 0, \sigma^2_{p_j}),\ \log(1-\pi_{p_j}) + \log \mathcal{N}(\beta_j \mid 0, \epsilon)\Big)$$

(Section S2).

For the covariate coefficients, an analogous marginalised mixture with a fixed inclusion probability $\pi_0$ is specified:

$$\gamma_k \sim \pi_0\,\mathcal{N}\big(0,\ \sigma^2_\gamma\big) + (1-\pi_0)\,\mathcal{N}(0,\ \epsilon)\,, \qquad k = 1, \ldots, K, \tag{S33}$$

$$\sigma^2_\gamma \sim \text{Inv-Gamma}(a_0, b_0). \tag{S34}$$

### S1.4.2 Marginal density

The hierarchy specifies the mixture marginals directly:

$$p(\beta_j) = \pi_{p_j}\,\mathcal{N}(\beta_j \mid 0, \sigma^2_{p_j}) + (1-\pi_{p_j})\,\mathcal{N}(\beta_j \mid 0, \epsilon), \tag{S35}$$

$$p(\gamma_k) = \pi_0\,\mathcal{N}(\gamma_k \mid 0, \sigma^2_\gamma) + (1-\pi_0)\,\mathcal{N}(\gamma_k \mid 0, \epsilon), \tag{S36}$$

which are exactly the densities evaluated by `log_mix` in the likelihood.

### S1.4.3 Posterior inclusion probability (PIP)

The PIP for $\beta_j$ is recovered analytically at each MCMC draw via Bayes' theorem:

$$\text{PIP}_j = \frac{\pi_{p_j}\,\phi(\beta_j;\,0, \sigma^2_{p_j})}{\pi_{p_j}\,\phi(\beta_j;\,0, \sigma^2_{p_j}) + (1-\pi_{p_j})\,\phi(\beta_j;\,0, \epsilon)}, \tag{S37}$$

where $\phi(\cdot;\,0, v)$ denotes the $\mathcal{N}(0, v)$ density. The posterior mean PIP, $\widehat{\text{PIP}}_j = \mathbb{E}[\text{PIP}_j \mid \mathcal{D}]$, is computed by averaging Equation (S37) over MCMC draws. A directed edge is declared active if $\widehat{\text{PIP}}_j \geq 0.5$.

### S1.4.4 Full conditionals

Because the mixture is marginalised over the component indicator, the slab-variance full conditional is obtained by introducing, for each draw, the posterior responsibility $r_j = \text{PIP}_j$ of the slab component (Equation (S37)). Conditioning on the responsibilities, the order-specific slab variances retain conjugate inverse-gamma updates,

$$\sigma^2_p \mid \boldsymbol{\beta}, \mathbf{r} \sim \text{Inv-Gamma}\left(a_0 + \tfrac{1}{2}\sum_{j:\,p_j=p} r_j,\ \ b_0 + \frac{1}{2}\sum_{j:\,p_j=p} r_j\,\beta_j^2\right), \tag{S38}$$

$$\sigma^2_\gamma \mid \boldsymbol{\gamma}, \mathbf{r}^\gamma \sim \text{Inv-Gamma}\left(a_0 + \tfrac{1}{2}\sum_{k} r^\gamma_k,\ \ b_0 + \frac{1}{2}\sum_{k} r^\gamma_k\,\gamma_k^2\right), \tag{S39}$$

with $r_k^\gamma$ the slab responsibility of $\gamma_k$. In the Stan implementation these auxiliary quantities are not instantiated: the mixture is evaluated directly through `log_mix` and all parameters $(\boldsymbol{\beta}, \boldsymbol{\gamma}, \boldsymbol{\sigma}^2)$ are sampled jointly by NUTS, with $\text{PIP}_j$ recomputed deterministically at each draw from Equation (S37).

## S1.5 Common Laplace-Approximation Form of the Full Conditional for $\beta_j$

Across all four priors the Poisson likelihood enters non-conjugately. A Laplace approximation around the current iterate yields the same approximate Gaussian form for the full conditional of $\beta_j$:

$$\log p(\beta_j \mid \text{rest}) \approx -\frac{(\beta_j - \mu_j)^2}{2\sigma_j^2} + \text{const}, \tag{S40}$$

with

$$\sigma_j^2 = \left( \frac{1}{v_j} + \sum_{i,s} x_{j,i,s}^2 \, T_{i,s} \, e^{\eta_{i,s}} \right)^{-1}, \qquad \mu_j = \sigma_j^2 \sum_{i,s} x_{j,i,s} \left( N_{i,s}^m - T_{i,s} \, e^{\eta_{i,s}^{(-j)}} \right), \tag{S41}$$

where $\eta_{i,s}^{(-j)}$ is the linear predictor excluding the contribution of $\beta_j$, and the effective prior variance $v_j$ differs by prior:

$$v_j = \begin{cases} \sigma_{p_j}^2 \, e^{-\theta p_j} & \text{(hierarchical structured)}, \\ 2\,\lambda_{p_j}^2 \, e^{-\theta p_j} & \text{(Bayesian LASSO; marginal Laplace variance)}, \\ \tilde{\lambda}_j^2 \, \tau_j^2 & \text{(regularised horseshoe)}. \end{cases}$$

For the spike-and-slab the full conditional of $\beta_j$ is instead a two-component Gaussian mixture; conditional on the slab component its effective variance is $\sigma_{p_j}^2$. This common form makes clear that the continuous priors differ only in how $v_j$ is determined from the auxiliary scale parameters. Exact posterior sampling is performed via NUTS in Stan.

## S1.6 Posterior Shrinkage Factor and Pseudo-PIP

This section derives the posterior shrinkage factor $\kappa_j$ used to construct the pseudo-PIPs for the three continuous priors (main text, Equation 3.3), and shows that it is genuinely per-coefficient.

Consider the Gaussian/Laplace approximation to the full conditional of $\beta_j$ derived above. The conditional posterior precision is the sum of the prior precision and the likelihood (Fisher) information,

$$\sigma_j^{-2} = v_j^{-1} + \mathcal{I}_j, \qquad \mathcal{I}_j = \sum_{i,s} x_{j,i,s}^2 \, T_{i,s} \, e^{\eta_{i,s}}, \tag{S42}$$

where $\mathcal{I}_j$ is the per-coefficient observed Fisher information for $\beta_j$ under the Poisson likelihood (the curvature $-\partial^2 \log p(\mathcal{D} \mid \boldsymbol{\theta}^m)/\partial \beta_j^2$ evaluated at the current iterate, equal to the $j$th

diagonal of $\mathbf{X}^\top \mathrm{diag}(\boldsymbol{\mu})\mathbf{X}$ with expected counts $\mu_{i,s} = T_{i,s} e^{\eta_{i,s}}$). The conditional posterior mean $\mu_j$ derived above is a precision-weighted average of the prior mean (0) and the data-only estimate $\hat{\beta}_j = \mathcal{I}_j^{-1} \sum_{i,s} x_{j,i,s}(N_{i,s}^m - T_{i,s} e^{\eta_{i,s}^{(-j)}})$:

$$\mu_j = \frac{v_j^{-1} \cdot 0 + \mathcal{I}_j \, \hat{\beta}_j}{v_j^{-1} + \mathcal{I}_j} = \frac{\mathcal{I}_j}{v_j^{-1} + \mathcal{I}_j} \, \hat{\beta}_j = (1 - \kappa_j) \, \hat{\beta}_j, \tag{S43}$$

so that the fraction of the posterior mean attributable to the prior (i.e. the amount of shrinkage toward zero) is

$$\kappa_j = \frac{v_j^{-1}}{v_j^{-1} + \mathcal{I}_j} = \frac{1}{1 + \mathcal{I}_j \, v_j} \qquad \in [0, 1]. \tag{S44}$$

Here $\kappa_j \to 1$ corresponds to complete shrinkage ($\beta_j$ pulled to zero) and $\kappa_j \to 0$ to a data-dominated, essentially unshrunk coefficient. Equation (S44) is the Poisson-GLM analogue of the shrinkage weight of Carvalho et al. (2010): in the canonical orthonormal-Gaussian regression with noise variance $\sigma^2$ and $n$ observations, $\mathcal{I}_j = n/\sigma^2$ for every $j$ and Equation (S44) reduces to their $\kappa_j = (1 + (n/\sigma^2) \, v_j)^{-1}$; the regularised-horseshoe special case $v_j = \tilde{\lambda}_j^2 \tau_j^2$ recovers $\kappa_j = (1 + (n/\sigma^2) \, \tau^2 \tilde{\lambda}_j^2)^{-1}$ of Piironen and Vehtari (2017).

Substituting the per-prior $v_j$ gives

$$\kappa_j = \begin{cases} \left(1 + \mathcal{I}_j \, \sigma_{p_j}^2 \, e^{-\theta p_j}\right)^{-1} & \text{(hierarchical structured)}, \\ \left(1 + 2\,\mathcal{I}_j \, \lambda_{p_j}^2 \, e^{-\theta p_j}\right)^{-1} & \text{(Bayesian LASSO)}, \\ \left(1 + \mathcal{I}_j \, \tilde{\lambda}_j^2 \, \tau_j^2\right)^{-1} & \text{(regularised horseshoe)}. \end{cases} \tag{S45}$$

The pseudo-PIP is $\widetilde{\mathrm{PIP}}_j = \mathbb{E}[1 - \kappa_j \mid \mathcal{D}]$, obtained by evaluating Equation (S44) at each posterior draw (using that draw's scale parameters and linear predictor, hence its $\mathcal{I}_j$) and averaging.

## S2 Stan Code for All Four Shrinkage Priors

The following R/Stan code implements the four shrinkage priors via the `rstan` package. Each continuous-prior model returns, in its `generated quantities` block, the per-coefficient shrinkage factor $\kappa_j$ of Section S1.6; the spike-and-slab returns the analytic PIP. Full reproducible code, including the data-preprocessing pipeline and the post-processing scripts that turn these quantities into pseudo-PIPs, PIPs, and evaluation metrics, is available at https://github.com/InflAIM-Study/CTBN-Multimorbidity-Paper.

```
# ============================================================================
# CTBN BAYESIAN POSTERIOR ESTIMATION  (rstan)
# Supports four prior specifications: spike_slab, structured, lasso, horseshoe.
# ============================================================================

# ---- 1. Spike-and-Slab (marginalised normal mixture; log_mix) --------------
.STAN_SPIKE_SLAB <- "
data {
  int<lower=1> N_rows;
  int<lower=1> Q;
  int<lower=1> P;
  array[N_rows] int<lower=0> N_obs;
  matrix[N_rows, Q] X;
  matrix[N_rows, P] Z;
  vector<lower=0>[N_rows] T;
  array[Q] int<lower=0> beta_order;
  real<lower=0, upper=1> pi0;
  real<lower=0>          theta;
  real<lower=0>          a0;
  real<lower=0>          b0;
  real<lower=0>          spike_var;
  int<lower=1>           max_order;
}
transformed data {
  vector[Q] pi_beta;
  for (j in 1:Q)
    pi_beta[j] = pi0 * exp(-theta * beta_order[j]);
  real pi_gamma = pi0;
}
parameters {
  vector[Q]                        beta;
  vector[P]                        gamma;
  vector<lower=0>[max_order + 1]   sigma2;
  real<lower=0>                    sigma2_gamma;
}
model {
  for (p in 0:max_order)
    sigma2[p + 1] ~ inv_gamma(a0, b0);
  sigma2_gamma ~ inv_gamma(a0, b0);

  for (j in 1:Q)
    target += log_mix(pi_beta[j],
                      normal_lpdf(beta[j]  | 0, sqrt(sigma2[beta_order[j] + 1])),
                      normal_lpdf(beta[j]  | 0, sqrt(spike_var)));
```

```
  for (k in 1:P)
    target += log_mix(pi_gamma,
                      normal_lpdf(gamma[k] | 0, sqrt(sigma2_gamma)),
                      normal_lpdf(gamma[k] | 0, sqrt(spike_var)));

  vector[N_rows] log_mu = X * beta + Z * gamma + log(T);
  target += poisson_log_lpmf(N_obs | log_mu);
}
generated quantities {
  vector[Q] pip_beta;
  vector[P] pip_gamma;
  for (j in 1:Q) {
    real ls = log(pi_beta[j]) + normal_lpdf(beta[j] | 0, sqrt(sigma2[beta_order[j]+1]));
    real lp = log1m(pi_beta[j]) + normal_lpdf(beta[j] | 0, sqrt(spike_var));
    pip_beta[j] = exp(ls - log_sum_exp(ls, lp));
  }
  for (k in 1:P) {
    real ls = log(pi_gamma) + normal_lpdf(gamma[k] | 0, sqrt(sigma2_gamma));
    real lp = log1m(pi_gamma) + normal_lpdf(gamma[k] | 0, sqrt(spike_var));
    pip_gamma[k] = exp(ls - log_sum_exp(ls, lp));
  }
}
"

# ---- 2. Hierarchical Structured -------------------------------------------
.STAN_STRUCTURED <- "
data {
  int<lower=1> N_rows;
  int<lower=1> Q;
  int<lower=1> P;
  array[N_rows] int<lower=0> N_obs;
  matrix[N_rows, Q] X;
  matrix[N_rows, P] Z;
  vector<lower=0>[N_rows] T;
  array[Q] int<lower=0> beta_order;
  real<lower=0> theta;
  real<lower=0> a0;
  real<lower=0> b0;
  int<lower=1>  max_order;
}
parameters {
  vector[Q]                      beta;
  vector[P]                      gamma;
  vector<lower=0>[max_order + 1] sigma2;
  real<lower=0>                  sigma2_gamma;
}
model {
  for (p in 0:max_order)
    sigma2[p + 1] ~ inv_gamma(a0, b0);
  sigma2_gamma ~ inv_gamma(a0, b0);

  for (j in 1:Q) {
    real eff_var = sigma2[beta_order[j] + 1] * exp(-theta * beta_order[j]);
```

```
    beta[j] ~ normal(0, sqrt(eff_var));
  }
  gamma ~ normal(0, sqrt(sigma2_gamma));
  vector[N_rows] log_mu = X * beta + Z * gamma + log(T);
  target += poisson_log_lpmf(N_obs | log_mu);
}
generated quantities {
  vector[Q] post_sd_beta;
  vector[Q] kappa;
  {
    // Per-coefficient Poisson Fisher information I_j = sum_r X[r,j]^2 * mu_r,
    // with mu_r = T_r * exp(eta_r) the expected count. The posterior shrinkage
    // factor kappa_j = 1 / (1 + I_j * v_j) is the fraction of the (Laplace-
    // approximate) conditional posterior mean attributable to the prior. It is
    // per-coefficient through I_j even though v_j is order-specific, so two
    // coefficients of the same interaction order with different data support no
    // longer share the same kappa. See Supplementary Section S1.6.
    vector[N_rows] mu_gq = T .* exp(X * beta + Z * gamma);
    for (j in 1:Q) {
      real v_j    = sigma2[beta_order[j] + 1] * exp(-theta * beta_order[j]);
      real info_j = dot_product(square(col(X, j)), mu_gq);
      post_sd_beta[j] = sqrt(v_j);
      kappa[j]        = 1.0 / (1.0 + info_j * v_j);
    }
  }
}
"

# ---- 3. Bayesian LASSO -----------------------------------------------------
.STAN_LASSO <- "
data {
  int<lower=1> N_rows;
  int<lower=1> Q;
  int<lower=1> P;
  array[N_rows] int<lower=0> N_obs;
  matrix[N_rows, Q] X;
  matrix[N_rows, P] Z;
  vector<lower=0>[N_rows] T;
  array[Q] int<lower=0> beta_order;
  real<lower=0> theta;
  real<lower=0> a0;
  real<lower=0> b0;
  int<lower=1>  max_order;
}
parameters {
  vector[Q]                       beta;
  vector[P]                       gamma;
  vector<lower=0>[max_order + 1] lambda2;
  real<lower=0>                   lambda2_gamma;
}
model {
  for (p in 0:max_order)
    lambda2[p + 1] ~ inv_gamma(a0, b0);
  lambda2_gamma ~ inv_gamma(a0, b0);
```

```
  for (j in 1:Q) {
    real scale_j = sqrt(lambda2[beta_order[j] + 1]) * exp(-theta * beta_order[j] / 2.0);
    beta[j] ~ double_exponential(0, scale_j);
  }
  gamma ~ double_exponential(0, sqrt(lambda2_gamma));

  vector[N_rows] log_mu = X * beta + Z * gamma + log(T);
  target += poisson_log_lpmf(N_obs | log_mu);
}
generated quantities {
  vector[Q] kappa;
  {
    // Per-coefficient Poisson Fisher information I_j = sum_r X[r,j]^2 * mu_r.
    // The Laplace marginal beta_j ~ DE(0, s_j) with s_j = sqrt(lambda2_p) *
    // exp(-theta*p/2) has prior variance v_j = 2*s_j^2 = 2*lambda2_p*exp(-theta*p).
    // Shrinkage factor kappa_j = 1 / (1 + I_j * v_j) is per-coefficient through
    // I_j. See Supplementary Section S1.6.
    vector[N_rows] mu_gq = T .* exp(X * beta + Z * gamma);
    for (j in 1:Q) {
      real v_j    = 2.0 * lambda2[beta_order[j] + 1] * exp(-theta * beta_order[j]);
      real info_j = dot_product(square(col(X, j)), mu_gq);
      kappa[j] = 1.0 / (1.0 + info_j * v_j);
    }
  }
}
"

# ---- 4. Regularised Horseshoe (Piironen & Vehtari 2017) -------------------
.STAN_HORSESHOE <- "
data {
  int<lower=1> N_rows;
  int<lower=1> Q;
  int<lower=1> P;
  array[N_rows] int<lower=0> N_obs;
  matrix[N_rows, Q] X;
  matrix[N_rows, P] Z;
  vector<lower=0>[N_rows] T;
  array[Q] int<lower=0> beta_order;
  real<lower=0> theta;
  real<lower=0> tau0;
  real<lower=0> slab_df;
  real<lower=0> slab_scale;
  int<lower=1>  max_order;
}
parameters {
  vector[Q]          z_beta;
  vector[P]          gamma;
  vector<lower=0>[Q] lambda;
  real<lower=0>      tau;
  real<lower=0>      c2;
}
transformed parameters {
  vector[Q] beta;
```

```
  {
    for (j in 1:Q) {
      real tau_j       = tau * exp(-theta * beta_order[j]);
      real lambda2_j   = square(lambda[j]);
      real lambda2_t   = c2 * lambda2_j / (c2 + square(tau_j) * lambda2_j);
      beta[j]          = z_beta[j] * sqrt(lambda2_t) * tau_j;
    }
  }
}
model {
  tau ~ cauchy(0, tau0);
  c2 ~ inv_gamma(0.5 * slab_df, 0.5 * slab_df * square(slab_scale));
  lambda ~ cauchy(0, 1);
  z_beta ~ std_normal();
  gamma ~ normal(0, 1);
  vector[N_rows] log_mu = X * beta + Z * gamma + log(T);
  target += poisson_log_lpmf(N_obs | log_mu);
}
generated quantities {
  vector[Q] kappa;
  {
    // Per-coefficient Poisson Fisher information I_j = sum_r X[r,j]^2 * mu_r,
    // replacing the homogeneous n/sigma^2 of the Gaussian case (Piironen &
    // Vehtari 2017) with the exposure-weighted information appropriate to the
    // Poisson GLM. v_j = lambda2_tilde_j * tau_j^2 is the (already local)
    // effective prior variance. See Supplementary Section S1.6.
    vector[N_rows] mu_gq = T .* exp(X * beta + Z * gamma);
    for (j in 1:Q) {
      real tau_j      = tau * exp(-theta * beta_order[j]);
      real lambda2_j  = square(lambda[j]);
      real lambda2_t  = c2 * lambda2_j / (c2 + square(tau_j) * lambda2_j);
      real v_j        = lambda2_t * square(tau_j);
      real info_j     = dot_product(square(col(X, j)), mu_gq);
      kappa[j] = 1.0 / (1.0 + info_j * v_j);
    }
  }
}
"
```

# S3 Detailed Prior Ranking Discussion

This section gives the detailed per-prior commentary supporting the consolidated ranking summary in the main text (Figure 6 of the main paper).

**Spike-and-slab.** The spike-and-slab ranks first across all three analysis models on RMSE, coverage distance from 0.95, FPR, and Poisson log-likelihood, demonstrating consistent superiority on parameter recovery, calibration, false-discovery control, and predictive accuracy simultaneously. Its sole and persistent weakness is TPR, on which it ranks last (4th) across all analysis models, reflecting the conservative nature of its hard-thresholding selection mechanism. Brier score ranking is 4th under A1, improving to 3rd under A2 and 2nd under A3.

Selection AUC ranks 3rd under A1 and A2, improving to 2nd under A3. Overall, the spike-and-slab presents the most balanced profile across estimation and selection metrics, with the trade-off of reduced sensitivity.

**Hierarchical Structured.** The structured normal ranks 2nd on RMSE across all analysis models, and achieves 1st on selection AUC under A2, indicating competitive PIP ranking ability under correct specification. However, it ranks poorly on coverage distance (2.5th under A1, 3rd under A2 and A3) and last on Brier score under A2 and A3 (4th), suggesting that while it recovers point estimates reasonably well, its uncertainty quantification and predictive calibration are weaker than the spike-and-slab. FPR ranks 3rd consistently, indicating moderate false-discovery control. The structured normal is thus a reasonable choice when point estimation is the priority but is not recommended when calibrated uncertainty or predictive performance is critical.

**Bayesian LASSO.** The Bayesian LASSO ranks last (4th) on RMSE across all analysis models and last on FPR, confirming poor parameter recovery and the weakest false-discovery control of all priors. Its ranks on TPR (2nd across all analysis models) and selection AUC (1st under A1 and A2, 3rd under A3) reflect its tendency to include most parameters, inflating both true and false positives simultaneously. Brier score is 1st under A1 and A2, suggesting competitive short-term predictive accuracy despite poor estimation, though this advantage disappears under A3 where its Brier rank falls to 3rd. The Bayesian LASSO is not recommended when parameter interpretability or network sparsity is required.

**Regularised horseshoe.** The regularised horseshoe ranks worst (4th) on coverage distance and selection AUC across all analysis models, consistent with its near-random pseudo-PIP discrimination documented in the main-text selection figures. It ranks 2nd on TPR across all analysis models, but as established earlier this reflects indiscriminate inclusion rather than genuine sensitivity. A notable exception is under A3, where the regularised horseshoe ranks 1st on coverage distance and 1st on Brier score, suggesting that its global shrinkage mechanism confers some advantage in uncertainty quantification and prediction when the analysis model is over-specified, though this does not compensate for its poor selection behaviour.

**Overall recommendation.** No single prior dominates uniformly across all seven metrics. However, the spike-and-slab presents the most coherent profile for the goals of network structure learning: it ranks first on RMSE, FPR, coverage calibration, and Poisson log-likelihood across all or most analysis models, at the cost of reduced sensitivity (TPR rank 4). The structured normal is a viable secondary choice when computational considerations favour a continuous prior, provided the analysis does not require hard variable selection. The Bayesian LASSO and regularised horseshoe are not recommended for this application given their systematic failure on FPR and selection AUC respectively.

# S4 Additional Predictive Performance Details

The under-specified analysis model A1 produces the largest predictive degradation: under A1, all priors yield Poisson log-likelihood of −0.5746 and Brier scores of 0.1581–0.1582, modestly worse than A2, reflecting the omission of true interaction terms from the linear predictor. Notably, under the over-specified analysis model A3, predictive performance is marginally better than or equivalent to A2 for most priors, with the spike-and-slab achieving the best Poisson log-likelihood (−0.5721) and joint-best Brier score (0.1572) across all analysis-model/prior combinations. The Bayesian LASSO is the sole exception, showing a deterioration under A3 relative to A2 (Poisson LL −0.5744 vs. −0.5731), consistent with its tendency to over-shrink or over-fit under an expanded parameter space.

The oracle efficiency analysis (Supplementary Figure S8) confirms that, across all three analysis models, the Brier-ratio distributions for all four priors are tightly concentrated at or just above 1.0 for most conditions, indicating that the fitted models approach oracle-level predictive efficiency regardless of prior choice or model specification. Conditions with greater baseline incidence heterogeneity across replicates (notably DM, DVL, HL, and OA) exhibit slightly wider Brier-ratio distributions, reflecting greater Monte Carlo variability in the oracle benchmark rather than genuine inefficiency of the fitted models. Under A3, the distributions remain tightly clustered near 1.0 for the spike-and-slab, structured normal, and regularised horseshoe, while the Bayesian LASSO shows a marginally wider spread in several conditions (e.g. DE, DVL). No prior exhibits systematic left-tail mass below 0.9 in any condition or analysis model, confirming that none of the fitted models suffer from meaningful predictive inefficiency relative to the oracle.

# S5 Supplementary Tables: Empirical Analysis

This section collects the detailed numerical tables underlying the application in Section 5 of the main text.

# S6 Supplementary Figures

This section collects supplementary figures referenced from the main text. Figures S1–S7 support the UK Biobank application (Section 5); Figure S8 supports the simulation study (Section 4). Further displays relocated from the main text to meet the page limit (Table S6 and Figures S9–S13) are collected in Section S7.

# S7 Additional Displays Relocated from the Main Text

This section collects displays relocated from the main text to meet the journal page limit. Table S6 gives the true covariate coefficients used in the simulation study (main-text Section 4). Figure S9 reports the threshold-free selection AUC underlying the simulation variable-selection results (Section 4). Figures S10 and S11 support the interaction-order

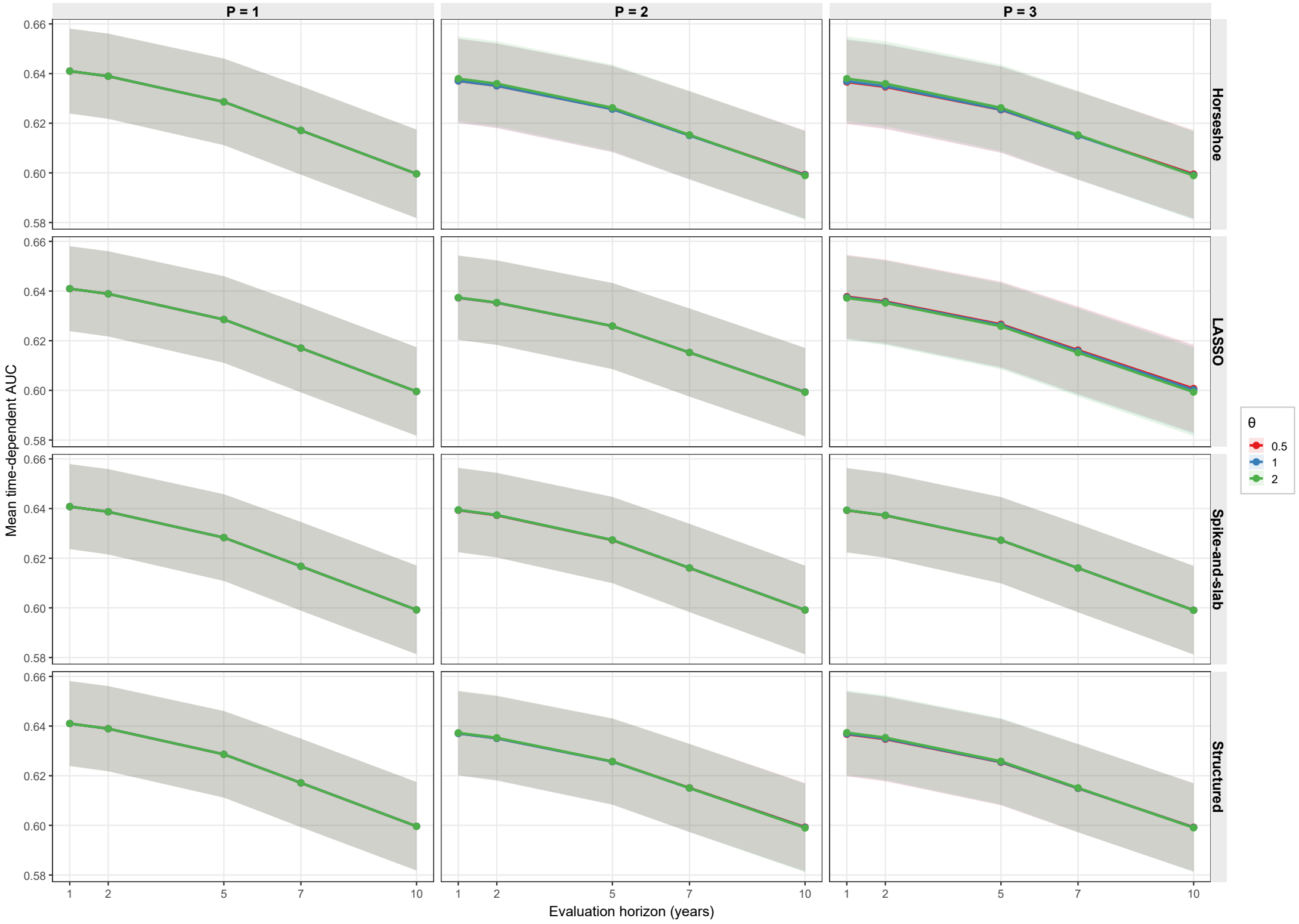


Figure S1: Sensitivity of the mean time-dependent AUC to the order penalty $\theta \in \{0.5, 1.0, 2.0\}$, faceted by prior family and interaction order $P$. Ribbons: $\pm 1.96\,\mathrm{SE}$ across folds.

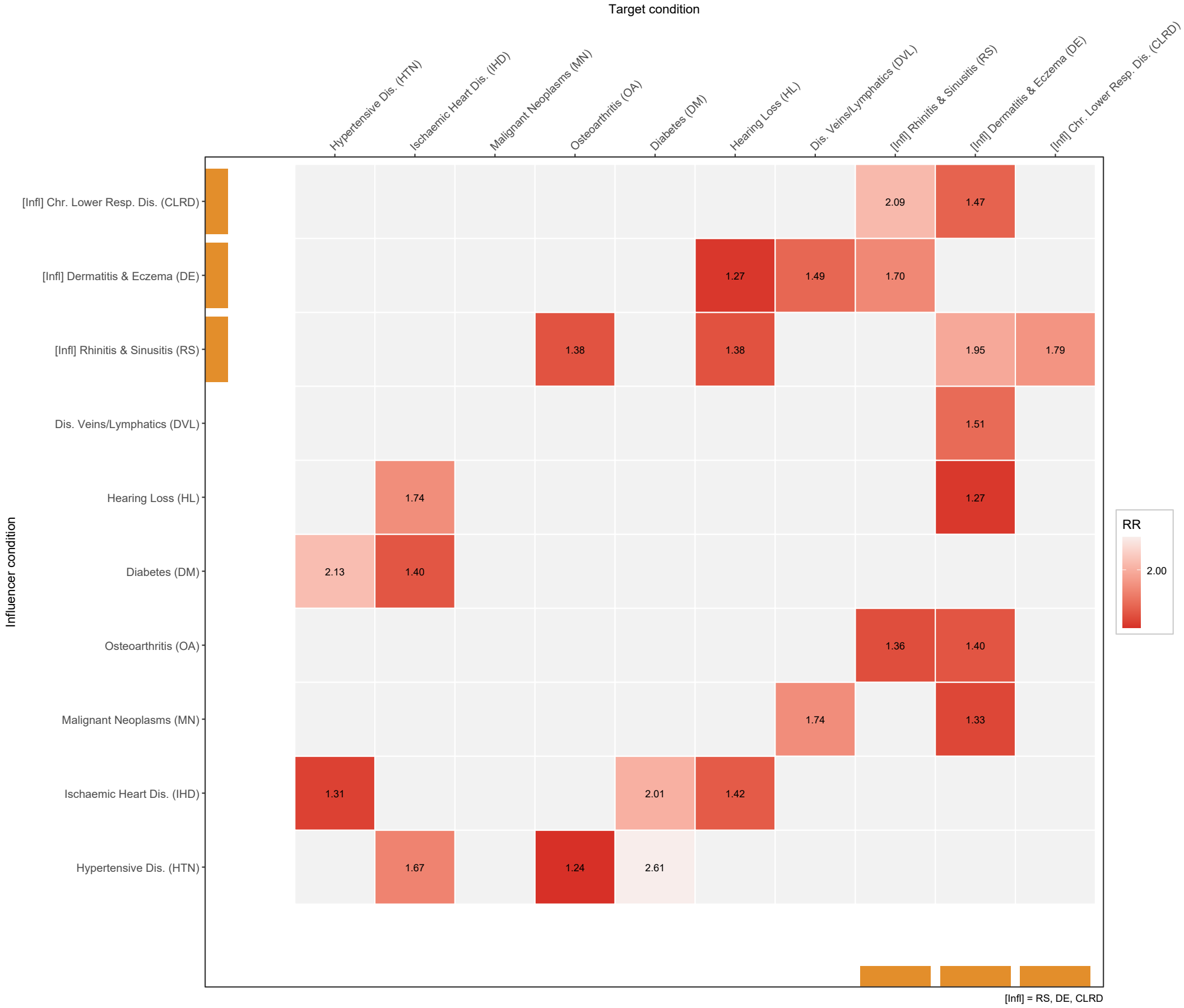


Figure S2: Posterior mean rate ratio for active network edges (PIP $\geq 0.50$). Blue, excitatory; red, inhibitory; grey, masked. The amber bar marks the inflammatory cluster.

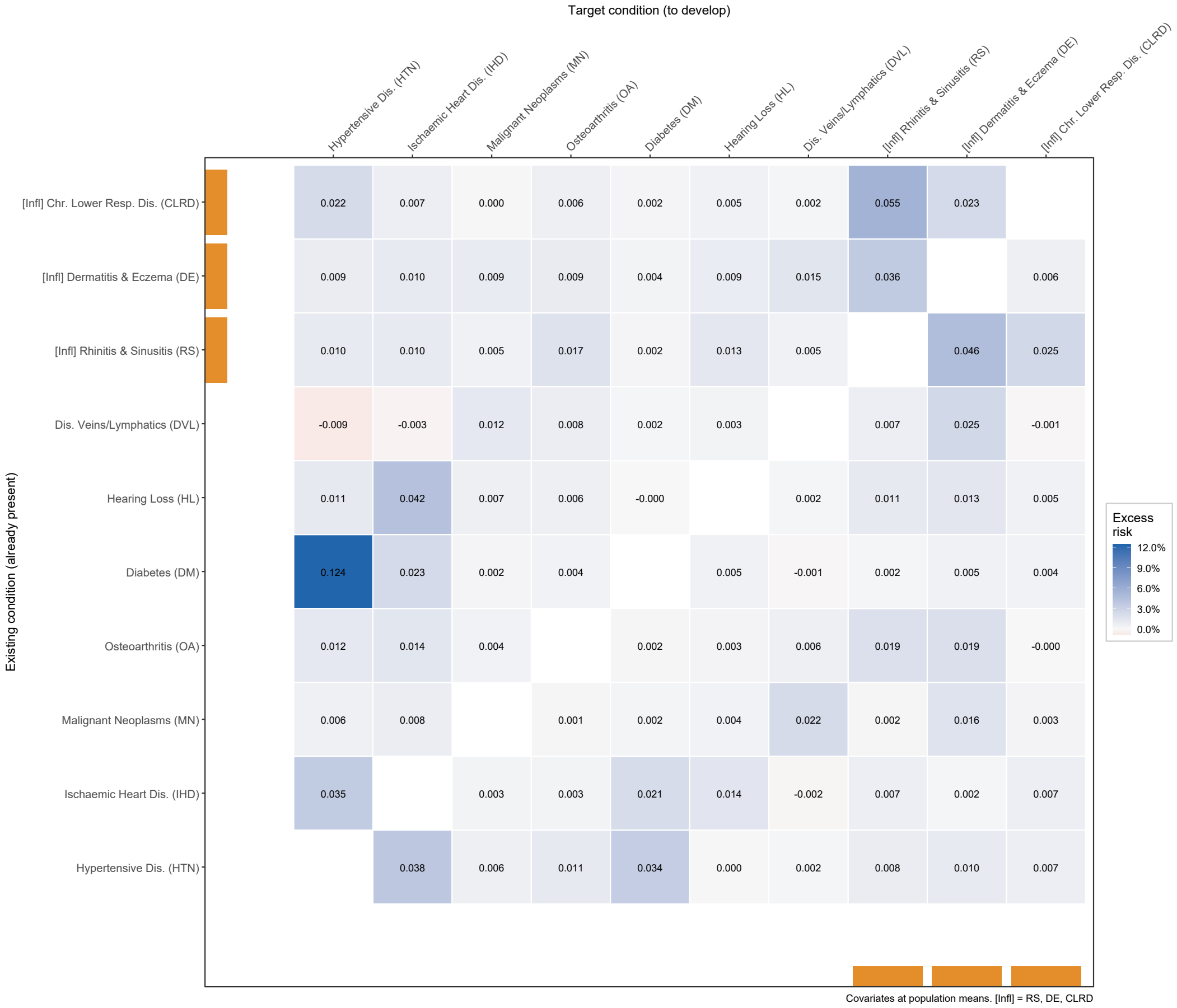


Figure S3: Five-year excess cumulative incidence $\Delta_m^{(1)}(j) = F_m(5 \mid X_j = 1, \mathbf{Z} = \bar{\mathbf{Z}}) - F_m(5 \mid X_j = 0, \mathbf{Z} = \bar{\mathbf{Z}})$ for all off-diagonal condition pairs. Blue, increased risk; red, reduced risk. The amber bar marks the inflammatory cluster; the diagonal is suppressed.

Table S2: Mean and standard error (SE) of the held-out Brier score and Poisson log-likelihood by prior family and interaction order $P$, for the UK Biobank application. Lower Brier score and higher (less negative) Poisson log-likelihood indicate better predictive performance.

| Prior | $P$ | Brier score mean (SE) | Poisson LL mean (SE) |
|---|---|---|---|
| Spike-and-Slab | 1 | 0.0635 (0.0025) | −0.2651 (0.0080) |
| | 2 | 0.0635 (0.0025) | −0.2651 (0.0080) |
| | 3 | 0.0635 (0.0025) | −0.2651 (0.0080) |
| Structured Normal | 1 | 0.0635 (0.0025) | −0.2652 (0.0080) |
| | 2 | 0.0635 (0.0025) | −0.2652 (0.0080) |
| | 3 | 0.0635 (0.0025) | −0.2652 (0.0080) |
| Bayesian LASSO | 1 | 0.0635 (0.0025) | −0.2652 (0.0080) |
| | 2 | 0.0635 (0.0025) | −0.2652 (0.0080) |
| | 3 | 0.0635 (0.0025) | −0.2654 (0.0080) |
| Reg. Horseshoe | 1 | 0.0635 (0.0025) | −0.2652 (0.0080) |
| | 2 | 0.0635 (0.0025) | −0.2652 (0.0080) |
| | 3 | 0.0635 (0.0025) | −0.2652 (0.0080) |

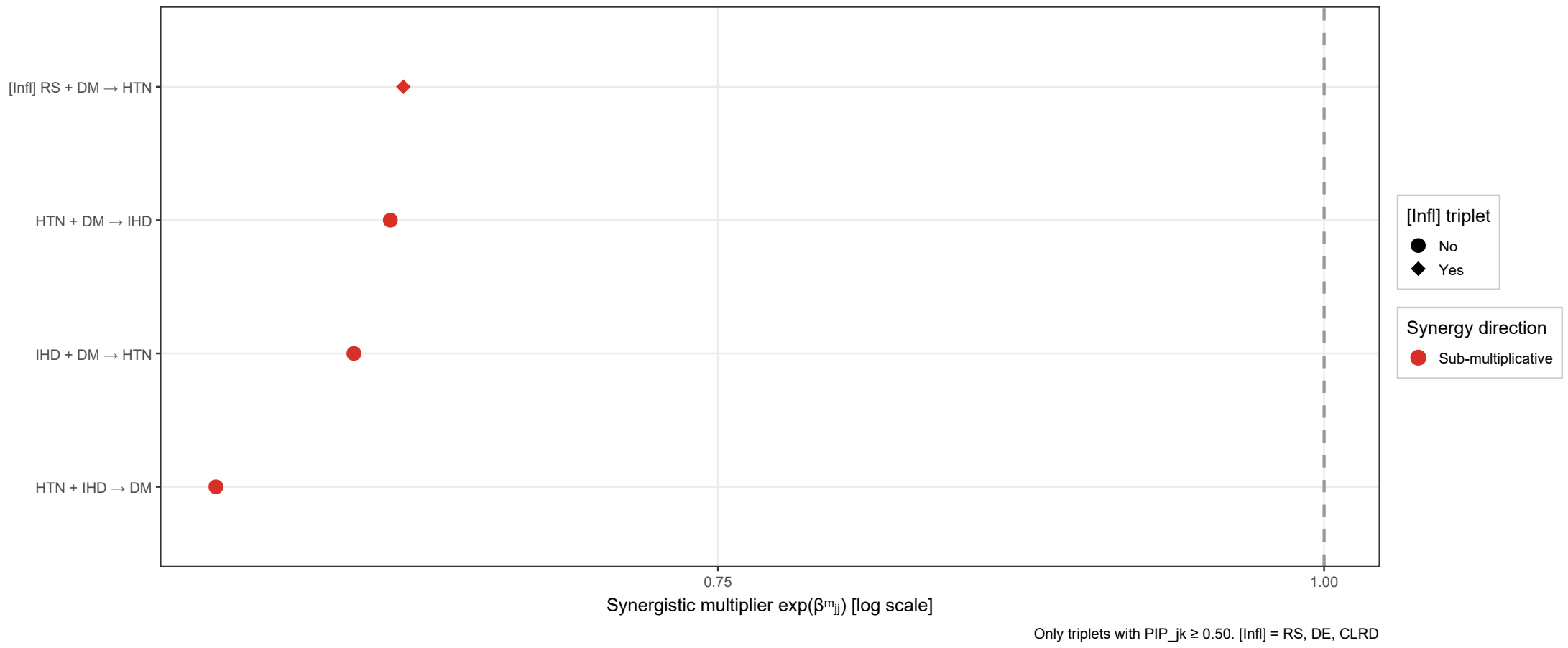


Figure S4: Synergistic multiplier $\exp(\hat{\beta}^m_{\{j,k\}})$ for all significant triplets ($\text{PIP}_{jk} \geq 0.50$). Blue ($> 1$), super-multiplicative; red ($< 1$), sub-multiplicative. A diamond ♦) marks triplets involving the inflammatory cluster.

Table S3: Active directed edges of the inferred multimorbidity network (PIP $\geq 0.50$), ordered by posterior inclusion probability. An asterisk (*) marks edges with at least one node in the inflammatory cluster.

| Influencer | Target | [Infl]* | PIP | log RR | RR | q0 (/1kpy) |
|---|---|---|---|---|---|---|
| HTN | IHD | | 1.000 | 0.5113 | 1.668 | 12.52 |
| HTN | DM | | 1.000 | 0.9593 | 2.610 | 4.34 |
| IHD | DM | | 1.000 | 0.6986 | 2.011 | 4.34 |
| [Infl] RS | [Infl] DE | * | 1.000 | 0.6655 | 1.945 | 10.46 |
| [Infl] RS | [Infl] CLRD | * | 1.000 | 0.5800 | 1.786 | 6.59 |
| [Infl] DE | [Infl] RS | * | 1.000 | 0.5319 | 1.702 | 11.02 |
| MN | DVL | | 1.000 | 0.5534 | 1.739 | 6.17 |
| [Infl] CLRD | [Infl] RS | * | 1.000 | 0.7352 | 2.086 | 11.02 |
| DM | HTN | | 1.000 | 0.7554 | 2.129 | 27.06 |
| HL | IHD | | 1.000 | 0.5543 | 1.741 | 12.52 |
| DVL | [Infl] DE | * | 0.998 | 0.4138 | 1.513 | 10.46 |
| [Infl] DE | DVL | * | 0.996 | 0.3988 | 1.490 | 6.17 |
| [Infl] CLRD | [Infl] DE | * | 0.993 | 0.3827 | 1.466 | 10.46 |
| IHD | HL | | 0.980 | 0.3539 | 1.425 | 6.89 |
| DM | IHD | | 0.965 | 0.3368 | 1.400 | 12.52 |
| OA | [Infl] DE | * | 0.960 | 0.3331 | 1.395 | 10.46 |
| [Infl] RS | OA | * | 0.949 | 0.3249 | 1.384 | 9.29 |
| [Infl] RS | HL | * | 0.944 | 0.3221 | 1.380 | 6.89 |
| OA | [Infl] RS | * | 0.915 | 0.3076 | 1.360 | 11.02 |
| MN | [Infl] DE | * | 0.849 | 0.2853 | 1.330 | 10.46 |
| IHD | HTN | | 0.777 | 0.2677 | 1.307 | 27.06 |
| HL | [Infl] DE | * | 0.628 | 0.2390 | 1.270 | 10.46 |
| [Infl] DE | HL | * | 0.623 | 0.2380 | 1.269 | 6.89 |
| HTN | OA | | 0.516 | 0.2187 | 1.244 | 9.29 |

Table S4: Top 15 condition pairs ranked by 5-year excess cumulative incidence. An asterisk (*) marks pairs involving at least one inflammatory condition.

| **Existing** | **Target** | **Infl.** | **Baseline (%)** | **Conditional (%)** | **Excess (%)** | **RR** |
|---|---|---|---|---|---|---|
| DM | HTN | | 12.6 | 25.0 | +12.4 | 2.13 |
| CLRD | RS | * | 5.4 | 10.9 | +5.5 | 2.09 |
| RS | DE | * | 5.1 | 9.7 | +4.6 | 1.95 |
| HL | IHD | | 6.1 | 10.3 | +4.2 | 1.74 |
| HTN | IHD | | 6.1 | 9.9 | +3.8 | 1.67 |
| DE | RS | * | 5.4 | 8.9 | +3.6 | 1.70 |
| IHD | HTN | | 12.6 | 16.2 | +3.5 | 1.31 |
| HTN | DM | | 2.1 | 5.5 | +3.4 | 2.61 |
| DVL | DE | * | 5.1 | 7.6 | +2.5 | 1.51 |
| RS | CLRD | * | 3.2 | 5.7 | +2.5 | 1.79 |
| DM | IHD | | 6.1 | 8.4 | +2.3 | 1.40 |
| CLRD | DE | * | 5.1 | 7.4 | +2.3 | 1.47 |
| CLRD | HTN | * | 12.6 | 14.9 | +2.2 | 1.19 |
| MN | DVL | | 3.0 | 5.2 | +2.2 | 1.74 |
| IHD | DM | | 2.1 | 4.3 | +2.1 | 2.01 |

Table S5: Pairwise interaction decomposition for the high-posterior-inclusion triplets ($\text{PIP}_{jk} \geq 0.50$; 5-year horizon). An asterisk (*) marks decompositions involving an inflammatory condition. Abbreviations: DM, diabetes; HTN, hypertensive diseases; IHD, ischaemic heart diseases; RS, rhinitis and sinusitis.

| **Target** | **Cond.** $j$ | **Cond.** $k$ | **Infl.** | **RR** $(j, k)$ | **Syn.** | **Joint RR** | $\boldsymbol{\Delta F_m(5)}$ | **PIP** |
|---|---|---|---|---|---|---|---|---|
| DM | HTN | IHD | | (2.61, 2.01) | 0.591 | 3.099 | ±1.1% | 1.000 |
| IHD | HTN | DM | | (1.67, 1.40) | 0.642 | 1.498 | ±3.3% | 0.998 |
| HTN | IHD | DM | | (1.31, 2.13) | 0.631 | 1.755 | ±7.4% | 0.999 |
| HTN | RS | DM | * | (1.08, 2.13) | 0.646 | 1.491 | ±7.7% | 0.997 |

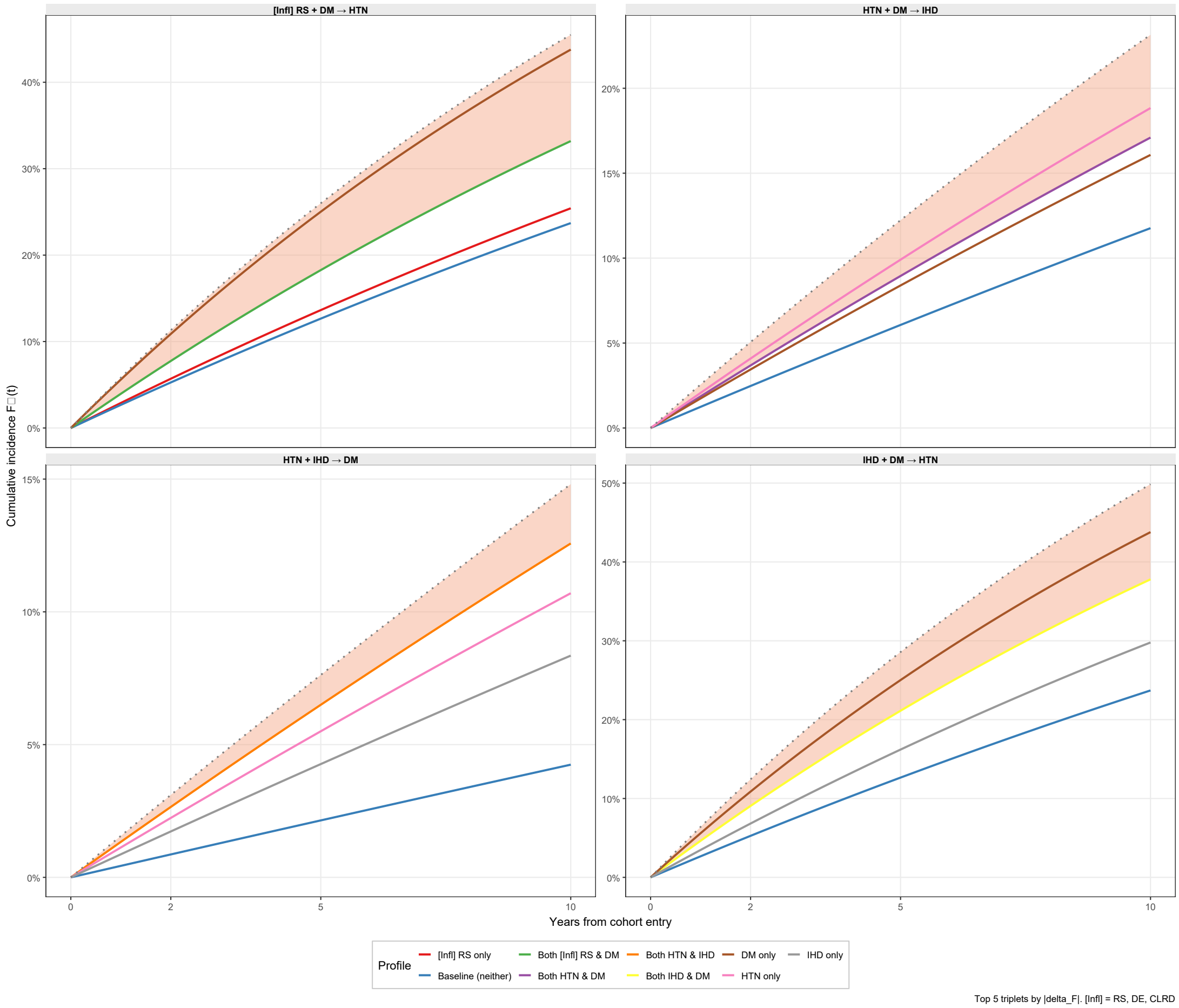


Figure S5: Cumulative incidence $\hat{F}_m(t)$ under four disease profiles for the four triplets with the largest significant $|\Delta F_m(5)|$. The orange ribbon marks the synergistic excess (joint minus additive counterfactual); the dotted line is the additive counterfactual. Covariates are held at population means.

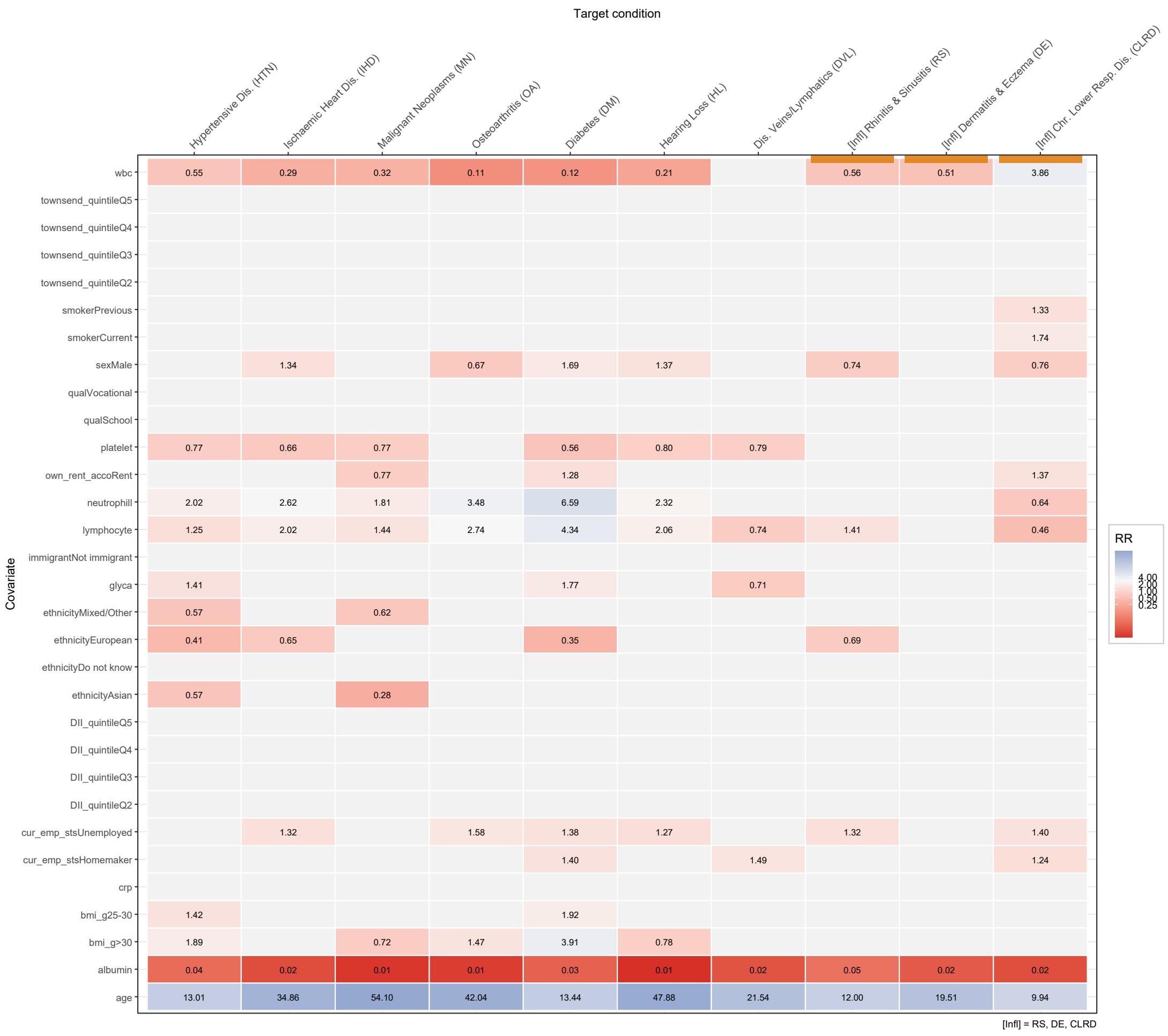


Figure S6: Posterior mean rate-ratio matrix for covariate–condition pairs. Grey cells have PIP $< 0.50$. The amber bar marks inflammatory target columns.

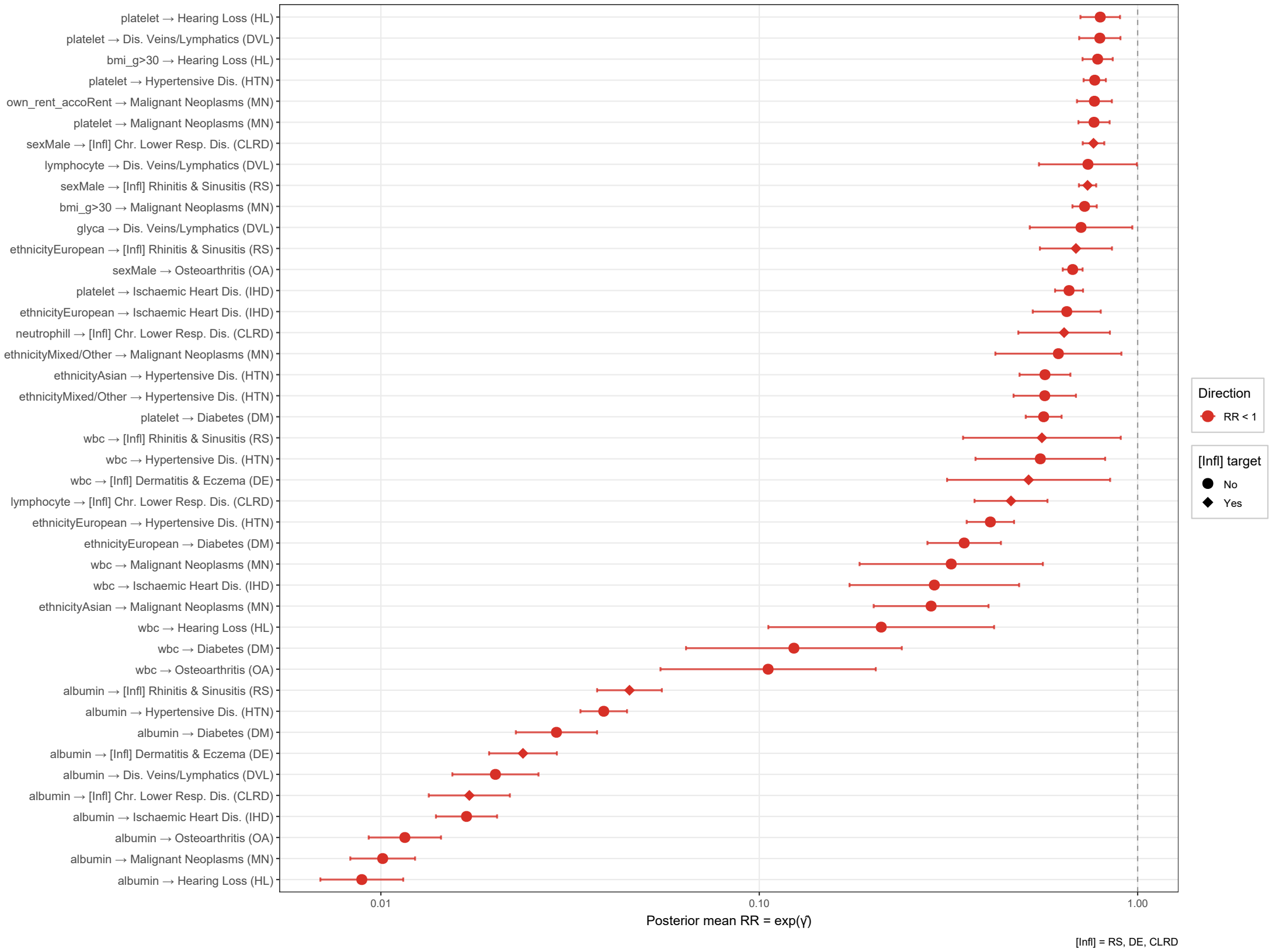


Figure S7: Forest plot of posterior mean rate ratios for inhibitory covariates (RR $< 1$ and PIP $\geq 0.50$), with 95% credible intervals from NUTS. Spike-and-slab prior; $\theta = 1$; $P = 2$.

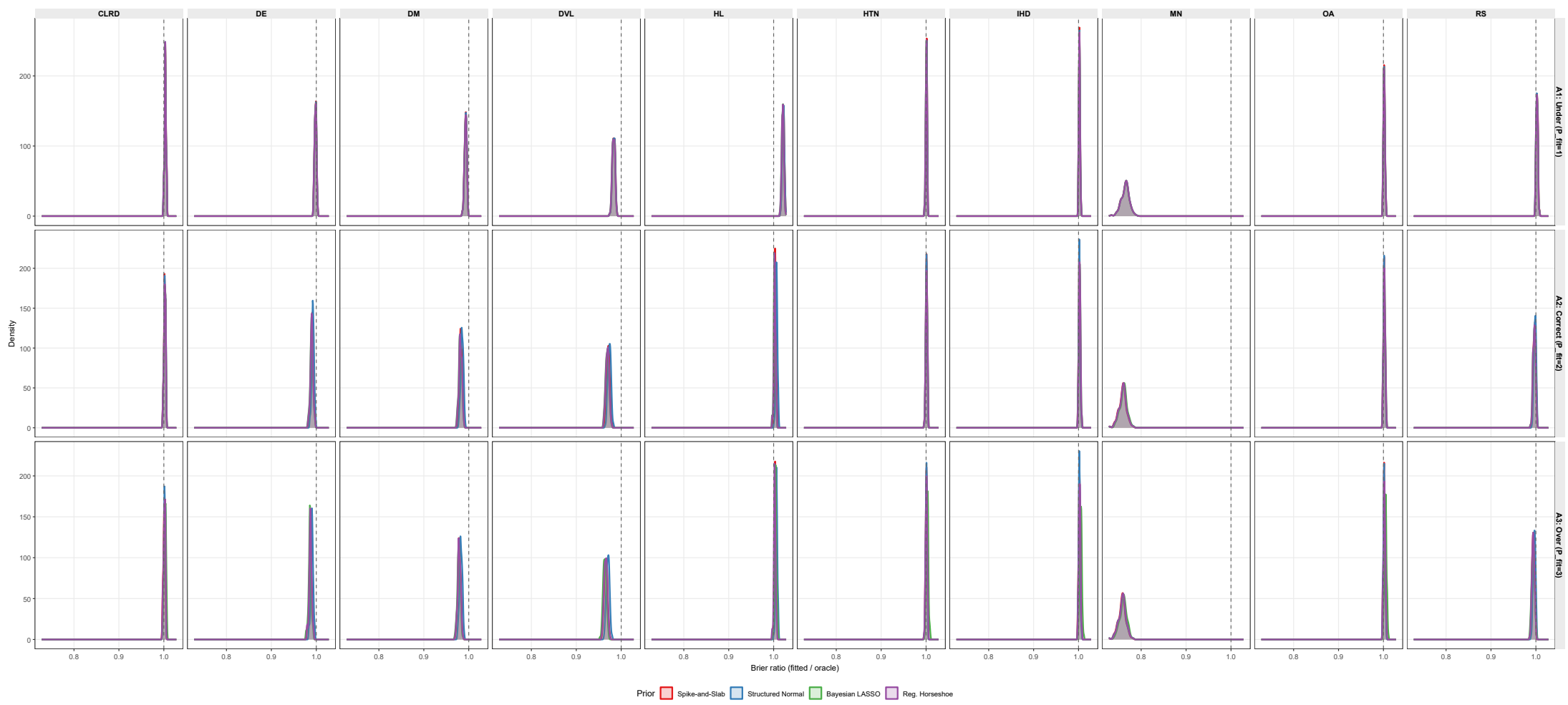


Figure S8: Oracle efficiency (fitted Brier / oracle Brier) by analysis model, condition, and prior. The dashed line marks 1.0; values above 1.0 indicate efficiency loss relative to the true-rate model.

and penalty sensitivity analysis of the application (Section 5), and Figure S12 gives the covariate posterior inclusion probabilities for the application (Section 5), and Figure S13 gives the consolidated prior-ranking heat-map for the simulation study (Section 4).

Table S6: True covariate coefficients ($\boldsymbol{\gamma}_m$) used in the simulation, calibrated to empirical estimates from the UK Biobank spike-and-slab analysis. Smoking effects are relative to the never-smoker reference category.

| **Condition** | **Age** | **Sex (male)** | **Smoking (current)** | **Smoking (former)** |
|---|---|---|---|---|
| Malignant neoplasms (MN) | 0.25 | 0.25 | 0.35 | 0.20 |
| Ischaemic heart diseases (IHD) | 0.30 | 0.25 | 0.40 | 0.25 |
| Dermatitis and eczema (DE) | 0.25 | −0.20 | 0.20 | 0.25 |
| Hypertensive diseases (HTN) | 0.35 | 0.20 | 0.25 | 0.20 |
| Diseases of veins/lymphatics (DVL) | 0.20 | 0.20 | 0.20 | 0.28 |
| Rhinitis and sinusitis (RS) | 0.20 | 0.25 | 0.25 | 0.25 |
| Hearing loss (HL) | 0.40 | 0.30 | 0.20 | 0.20 |
| Diabetes mellitus (DM) | 0.25 | 0.25 | 0.30 | 0.22 |
| Osteoarthritis (OA) | 0.45 | 0.20 | 0.25 | 0.25 |
| Chronic lower respiratory diseases (CLRD) | 0.30 | 0.20 | 0.60 | 0.20 |

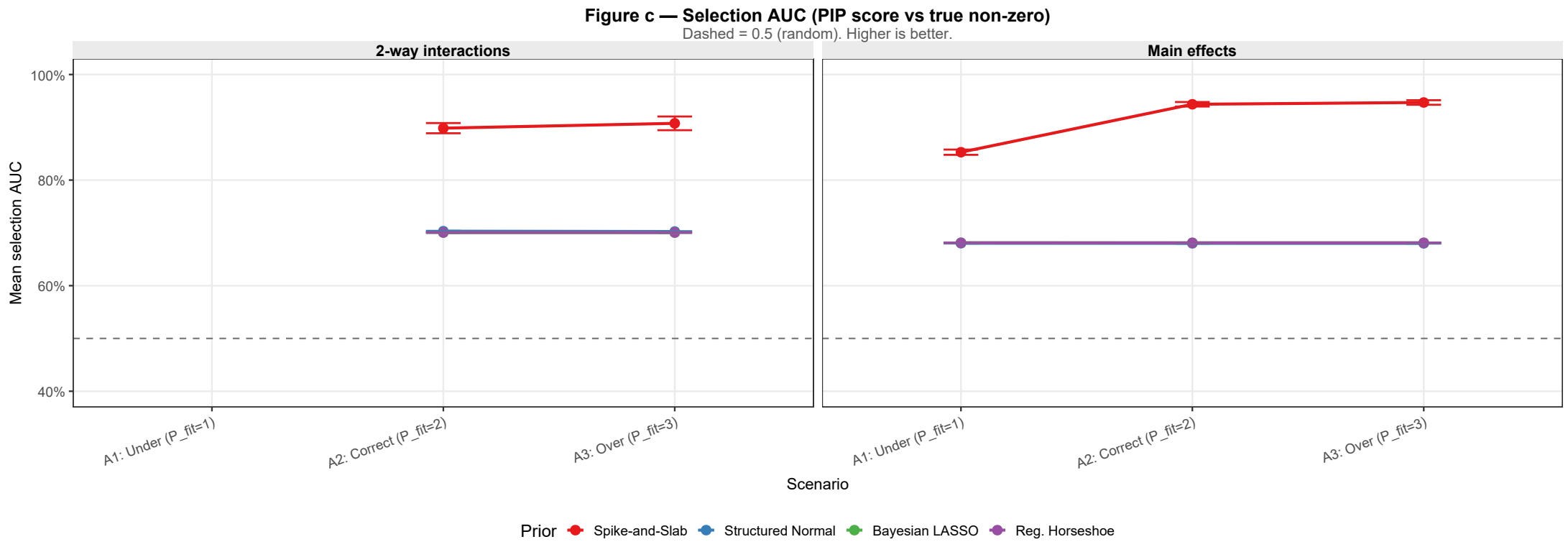


Figure S9: Selection AUC by analysis model, coefficient type, and prior (simulation study, main-text Section 4). Dashed line at AUC $= 0.50$ (random discrimination).

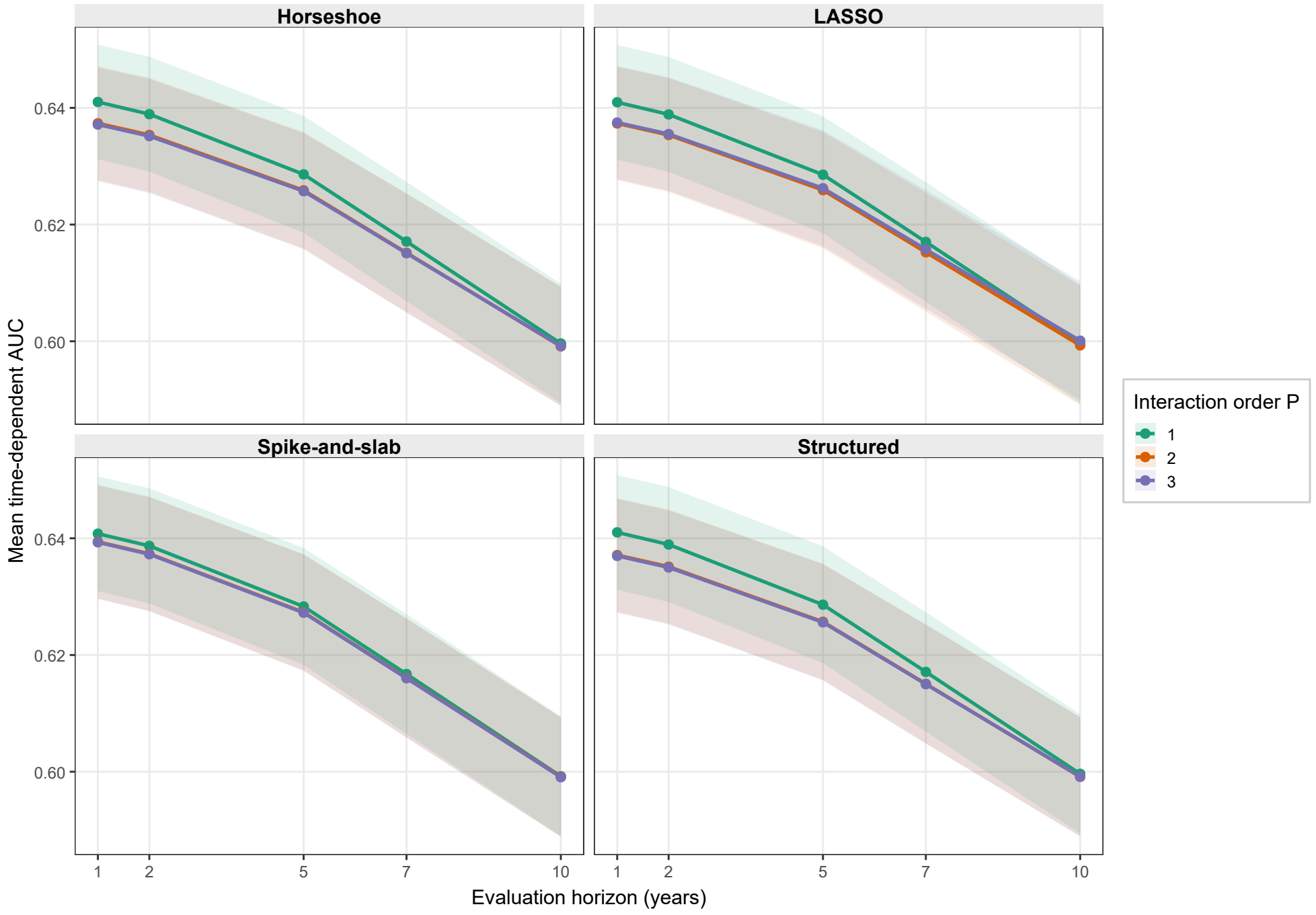


Figure S10: Time-dependent AUC by interaction order $P \in \{1, 2, 3\}$ across all prior families (application, main-text Section 5). Evaluation horizons $\tau \in \{1, 2, 5, 7, 10\}$ years are shown as labelled tick marks on the x-axis. Ribbons: $\pm 1.96\,\mathrm{SE}$ across folds.

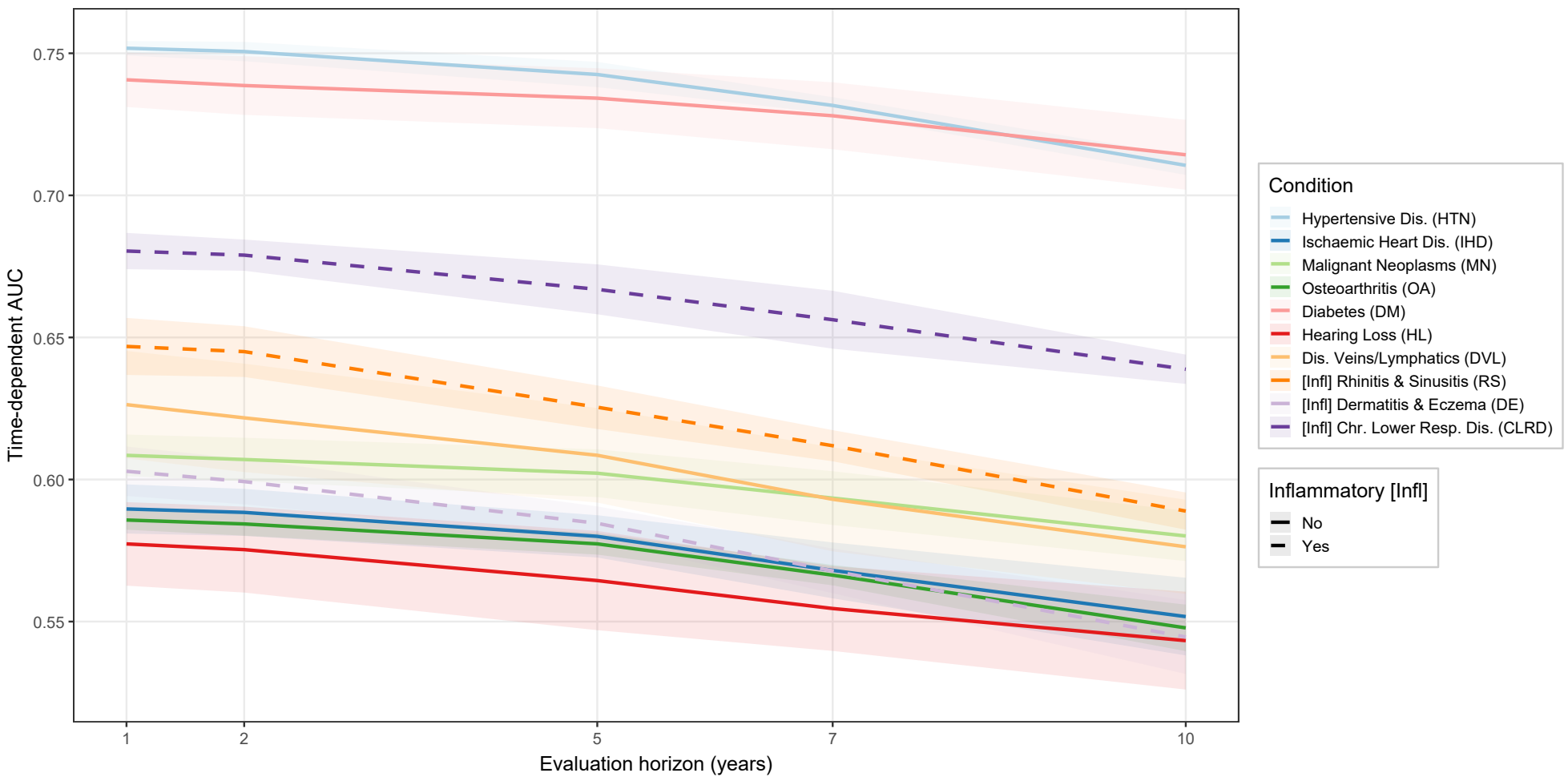


Figure S11: Per-condition tdAUC for the recommended model (spike-and-slab, $P = 2$, $\theta = 1$; application, main-text Section 5). **Dashed lines** = inflammatory conditions. Ribbons: $\pm 1.96\,\mathrm{SE}$ across folds.

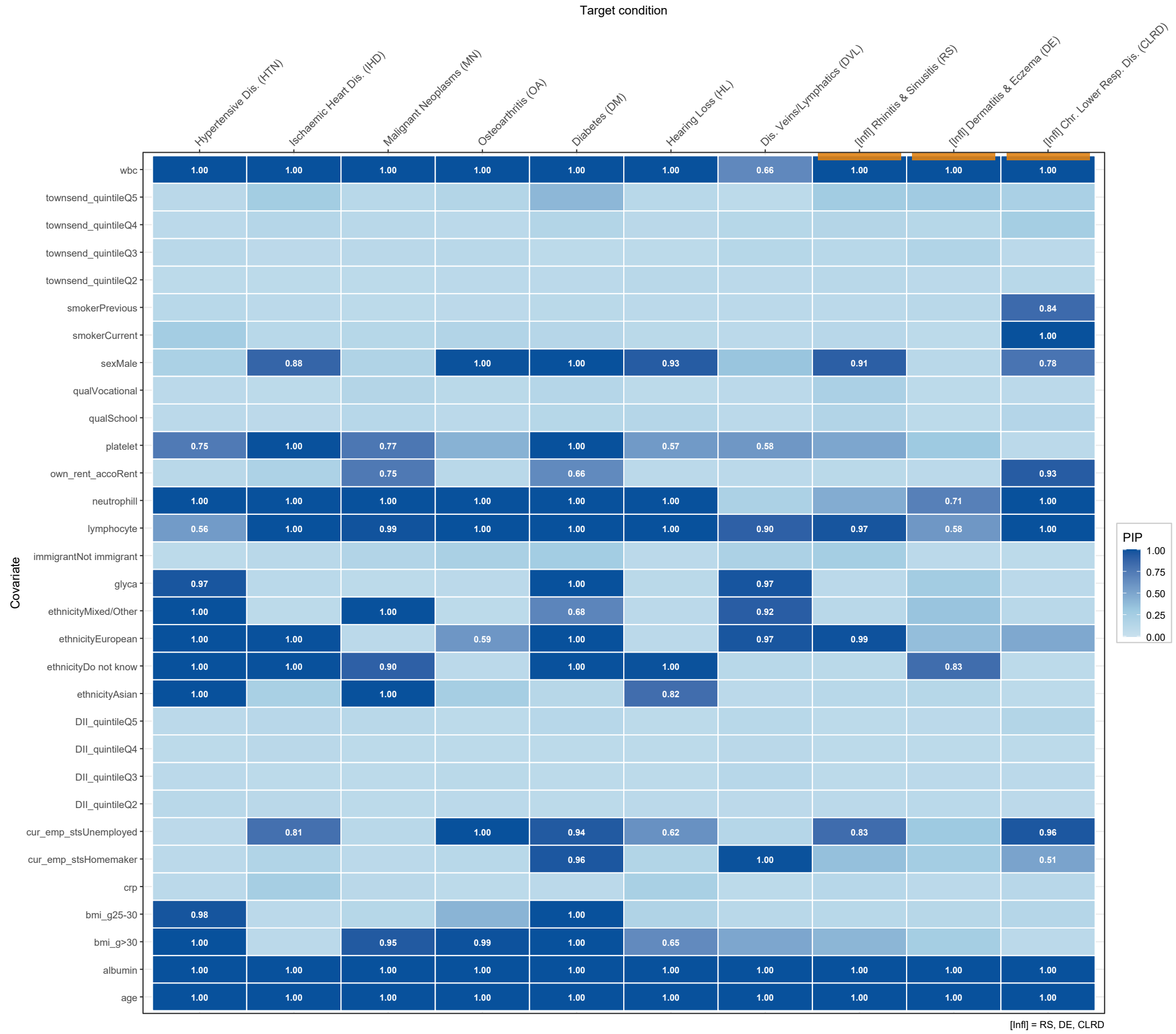


Figure S12: Posterior inclusion probabilities for covariate effects (application, main-text Section 5). Bold white text: PIP $\geq 0.50$. Amber bar = inflammatory target columns. Spike-and-slab; $\theta = 1$; $P = 2$.

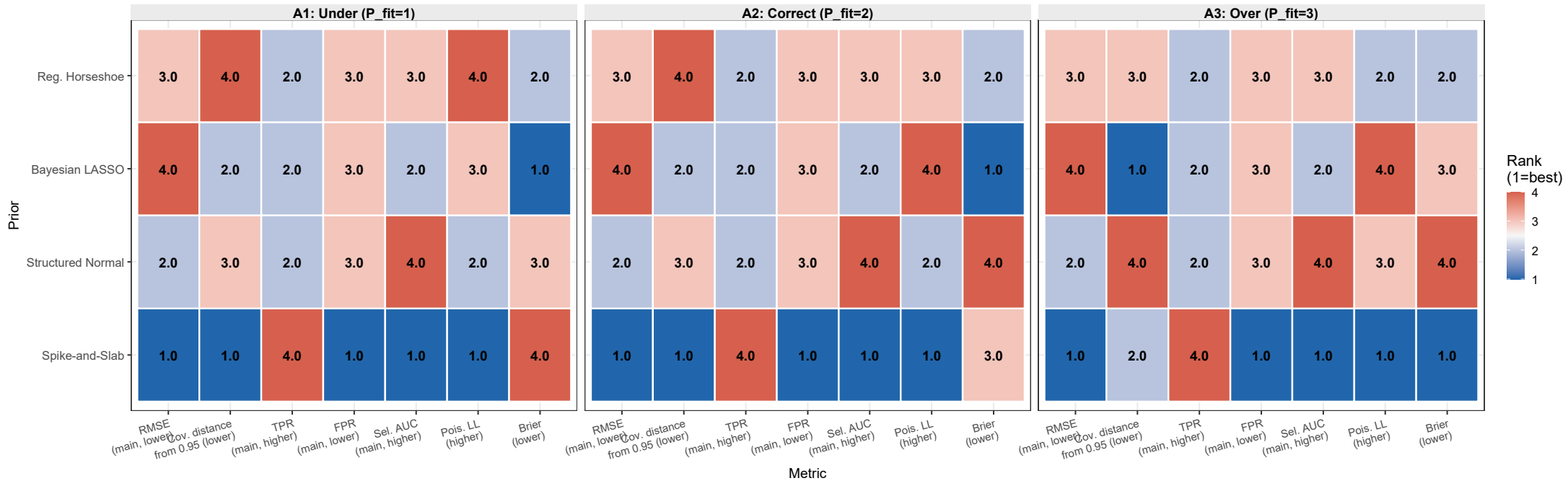


Figure S13: Prior ranking heat-map by analysis model and metric (simulation study, main-text Section 4). Rank 1 (blue) = best; Rank 4 (red) = worst. Metrics: RMSE, coverage deviation from 0.95, TPR, FPR, selection AUC, Poisson log-likelihood, Brier score.


\end{document}